\documentclass[aps,twocolumn,showpacs]{revtex4}
\makeatletter

\newcommand{\Rmnum}[1]{\expandafter\@slowromancap\romannumeral #1@}
\makeatother
\usepackage{graphicx}
\usepackage{dcolumn}
\usepackage{bm}
\usepackage{CJK}
\usepackage{color}
\usepackage{amsfonts}
\usepackage{psfrag}
\usepackage{wrapfig}
\usepackage{subfigure}
\usepackage{makeidx}
\usepackage{multirow}
\usepackage{epsf}
\usepackage{hyperref}
\usepackage{amsmath}
\usepackage{cases}

\begin{document}

\title{Femtosecond optical superregular breathers}
\author{Chong Liu$^{1,2}$}\email{chongliu@nwu.edu.cn}
\author{Lei Wang$^{3}$}\email{50901924@ncepu.edu.cn}
\author{Zhan-Ying Yang$^{1,2}$}\email{zyyang@nwu.edu.cn}
\author{Wen-Li Yang$^{2,4}$}
\address{$^1$School of Physics, Northwest University, Xi'an 710069, P.\ R.\ China}
\address{$^2$Shaanxi Key Laboratory for Theoretical Physics Frontiers, Xi'an 710069, P.\ R.\ China}
\address{$^3$Department of Mathematics and Physics, North China Electric Power University, Beijing 102206, P.\ R.\ China}
\address{$^4$Institute of Modern Physics, Northwest University, Xi'an 710069, P.\ R.\ China}
\begin{abstract}
Superregular (SR) breathers are nonlinear wave structures formed by a unique nonlinear
superposition of pairs of quasi-Akhmediev breathers. They describe a complete scenario of modulation instability that develops from localized small perturbations as well as an unusual quasiannihilation of breather collision.
Here, we demonstrate that femtosecond optical SR breathers in optical fibers exhibit intriguing half-transition and full-suppression states, which are absent in the picosecond regime governed by the standard nonlinear Schr\"{o}dinger equation. 
In particular, the full-suppression mode, which is strictly associated with the regime of vanishing growth rate of modulation instability, reveals a crucial \textit{non-amplifying} nonlinear dynamics of localized small perturbations. We numerically confirm the robustness of such different SR modes excited from ideal and nonideal initial states in both integrable and nonintegrable cases.
\end{abstract}

\maketitle
\section{Introduction}
Modulation instability (MI) is a nonlinear phenomenon of fundamental importance in many physical contexts ranging from hydrodynamics, nonlinear optics, plasma,
and ultracold atoms \cite{MI0,MI1}. In particular, MI is regarded as the origin of solitons, supercontinuum generation \cite{MI2}, and rogue wave events \cite{MIr0,MIr1,MIr2,MIr3,MIr4,MIr5}.
Whenever MI is induced by perturbations, the resulting process consists of an initial \textit{linear} stage and a subsequent \textit{nonlinear} stage.
For the former, the spectral sidebands associated with the instability
experience exponential amplification at the expense of the pump, but the latter exhibits richer dynamics.

An elementary prototype of the complete MI dynamics evolved from a weak periodic
perturbation on a continuous wave background is known as Akhmediev breathers \cite{AB1,AB2}.
Such breather, however, has its own limitations from the perspective of physical and practical significance. Namely, this MI scenario exhibits only one growth-return cycle and requires a specially constructed initial periodic perturbation in the whole infinite space. Nevertheless, these restrictions can be overcome when one goes beyond the elementary instability by considering the superregular (SR) breather \cite{SR1,SR2,SR3}. The latter 
is formed by a nonlinear superposition of multiple quasi-Akhmediev breathers. In this regard, the SR breather scenario of MI is associated with higher-order MI \cite{MIho} exhibiting complex evolution of multiple quasi-Akhmediev modes, but is developed from a localized small perturbation at the central position of mode interaction \cite{SR1,SR2,SR3}.
On the other hand, the reverse process describes a unique quasiannihilation of multiple breathers \cite{SR1,SR2,SR3}, which is absent in the traditional soliton interaction theory \cite{SI1,SI2}.
Interestingly, this complete SR breather has been
observed in the picosecond regime of optical fiber \cite{SR3}, ruled by the standard nonlinear Schr\"{o}dinger equation (NLSE). One should note that the SR mode is not the only complete MI scenario with complex nonlinear stage.
In fact, significant progresses have been recently made on different scenarios of the nonlinear development of MI, including
the role of continuous spectrum \cite{Biondini}, integrable turbulence \cite{IT1,IT2,IT3}, Fermi-Pasta-Ulam recurrence \cite{FPU1,FPU2},
and heteroclinic mode \cite{Conforti}. The common features and differences among their dynamics manifestations enrich the MI understanding of nonlinear stage.

On the other hand,
a linear stability analysis is generally performed to study the initial linear stage of MI. It provides, nevertheless, a straightforward way to identify the instability criterion
and to evaluate the growth rate of small perturbations. In particular, recent studies have been devoted to the crucial links between two complete MI scenarios (rogue wave generation \cite{MIr2} and thermalization of Fermi-Pasta-Ulam recurrence \cite{FPU1}) and the linearized results, although a comprehensive investigation of exact relations between various MI scenarios and the linear stability analysis still remains largely unexplored.


In this work, we study the existence, characteristic, and mechanism of optical SR modes in the femtosecond regime of optical fiber.
However, in this case higher-order effects such as higher-order dispersion, self-steepening, and self-frequency shift come into play and
impact strongly the instability criterion and growth rate distribution of MI \cite{hMI1,hMI3,hMI4,hMI5} by making optical fiber systems convectively unstable
\cite{Mussot}. As a result, femtosecond nonlinear modes on a continuous wave background can exhibit structural diversity beyond the reach of the standard NLSE \cite{n1,n2,n3,n4,n5,n6,n7,n8,n9,n10}. Here we unveil, analytically and numerically,
two intriguing states of femtosecond SR waves (i.e., the \textit{half-transition} and \textit{full-suppression} modes), arising from the higher-order effects. Our results demonstrate that these different SR modes can be evolved from an identical localized small perturbation.
In particular, we numerically confirm the robustness of such different SR modes evolution from the ideal and nonideal initial states in both the integrable and nonintegrable cases.

\section{Model and fundamental excitations}
Femtosecond pulses (i.e., the duration is shorter than 100 fs) propagation in optical fibers with higher-order physical effects is governed by the following higher-order NLSE \cite{b1,b2}
\begin{eqnarray}
\label{equ1}
i \frac{\partial u}{\partial z}+\frac{1}{2}\frac{\partial^2u}{\partial t^2}+|u|^2u-i\beta \frac{\partial^3u}{\partial t^3}-is\frac{\partial|u|^2u}{\partial t}\nonumber\\
-(i\gamma+t_R)u\frac{\partial|u|^2}{\partial t}=0,
\end{eqnarray}
where $u$ is the slowly varying envelope of the electric field, $z$ is the propagation distance, $t$ is the retarded time.
Parameters $\beta$, $s$, $\gamma$, $t_R$ are real constants which are responsible for the third-order dispersion, self-steepening, nonlinear dispersion, and self-frequency shift effects, respectively. In this general case, Eq. (\ref{equ1}) is not integrable.
To study femtosecond SR waves exactly, we shall omit the self-frequency shift effect, i.e., $t_R=0$, and consider a special condition for the higher-order terms, i.e., $s=6\beta$, $s+\gamma=0$.
As a result, the model (\ref{equ1}) reduces to the well-known Hirota equation \cite{Hirota}.
The latter has been demonstrated to be one of nontrivial integrable generalizations of the standard NLSE, since it admits
some interesting nonlinear modes which are absent in the standard NLSE \cite{n6,n7,Demontis}. However, the unique characteristic of SR modes in Eq. (\ref{equ1}) remain unexplored so far.

One should note that although the exact SR mode solutions are constructed in the specific integrable case, the robustness of SR modes excited from ideal and nonideal initial states in both integrable and nonintegrable cases will be confirmed by the direct numerical simulations (see below). The analytical and numerical results could give a generalized picture of the femtosecond SR wave features with higher-order effects.


Since SR modes are formed by the nonlinear superposition of multi-fundamental excitations, we shall first shed light on the properties of the fundamental modes in the femtosecond regime.
To this end, we construct first-order solution via the Darboux transformation \cite{DT1} with the spectral parameter $\lambda$ parameterized by Jukowsky transform \cite{SR1,SR2,SR3}.
The analytical first-order fundamental solution with a general and concise form is given by
\begin{equation}
u=u_0\left[1-4\rho\frac{\cosh{(\Theta+i\alpha)-\cos{(\Phi-\psi)}}}
{r\cosh{\Theta}-2\cos{\alpha}\cos{\Phi}}\right],\label{equs}
\end{equation}
where $u_0=ae^{i\theta}$, $\theta=q t+[a^2- q^2/2+\beta(6qa^2-q^3)]z$ denotes the initial continuous wave background with the amplitude $a$ and frequency $q$.
The perturbation on the background $u_0$, formed by a nonlinear superposition of the hyperbolic function $\cosh{\Theta}$ and trigonometric function $\cos{\Phi}$, exhibits unique characteristics of fundamental excitations. The corresponding explicit expressions read:
\begin{gather}
\Theta=2\eta_r(t-V_{gr}z)+\mu_1,~~\Phi=2\eta_i(t-V_{ph}z)-\theta_1,  \\
V_{gr}=v_1+v_2\frac{\eta_i}{\eta_r}, ~~V_{ph}=v_1-v_2\frac{\eta_r}{\eta_i}, \\
v_1=q+\varrho_1-\beta(2a^2-3q^2+4\rho^2-6q\varrho_1-4\varrho_1^2),\\
v_2=-6\beta q_{s1}\rho(1-q/q_{s1}),~q_{s1}=-\frac{1}{6\beta}(1+8\varrho_1\beta),\\
\eta_r=\frac{a}{2}\left(R-1/R\right)\cos{\alpha},~\eta_i=\frac{a}{2}\left(R+1/R\right)\sin{\alpha}, \\
\varrho_1=\frac{a}{2}\left(R-1/R\right)\sin{\alpha},~\rho=\frac{a}{2}\left(R+1/R\right)\cos{\alpha},  \\
r=a(R+1/R),~~\psi=\arctan[(i-iR^2)/(1+R^2)].
\label{eq1}
\end{gather}

The above expressions depend on the background amplitude $a$, frequency $q$,
the phase parameters $\theta_1$, $\mu_1$, and the real parameters $R$, $\alpha$. Note that $R$ $(\geq1)$, $\alpha$ are the radius and angle of the polar coordinates, which are introduced by the Jukowsky transform of the spectral parameter, i.e., $\lambda=i\frac{a}{2}(\xi+\frac{1}{\xi})-\frac{q}{2}$ where $\xi=Re^{i\alpha}$.
\begin{figure}[htb]
\centering
\subfigure{\includegraphics[height=42mm,width=42mm]{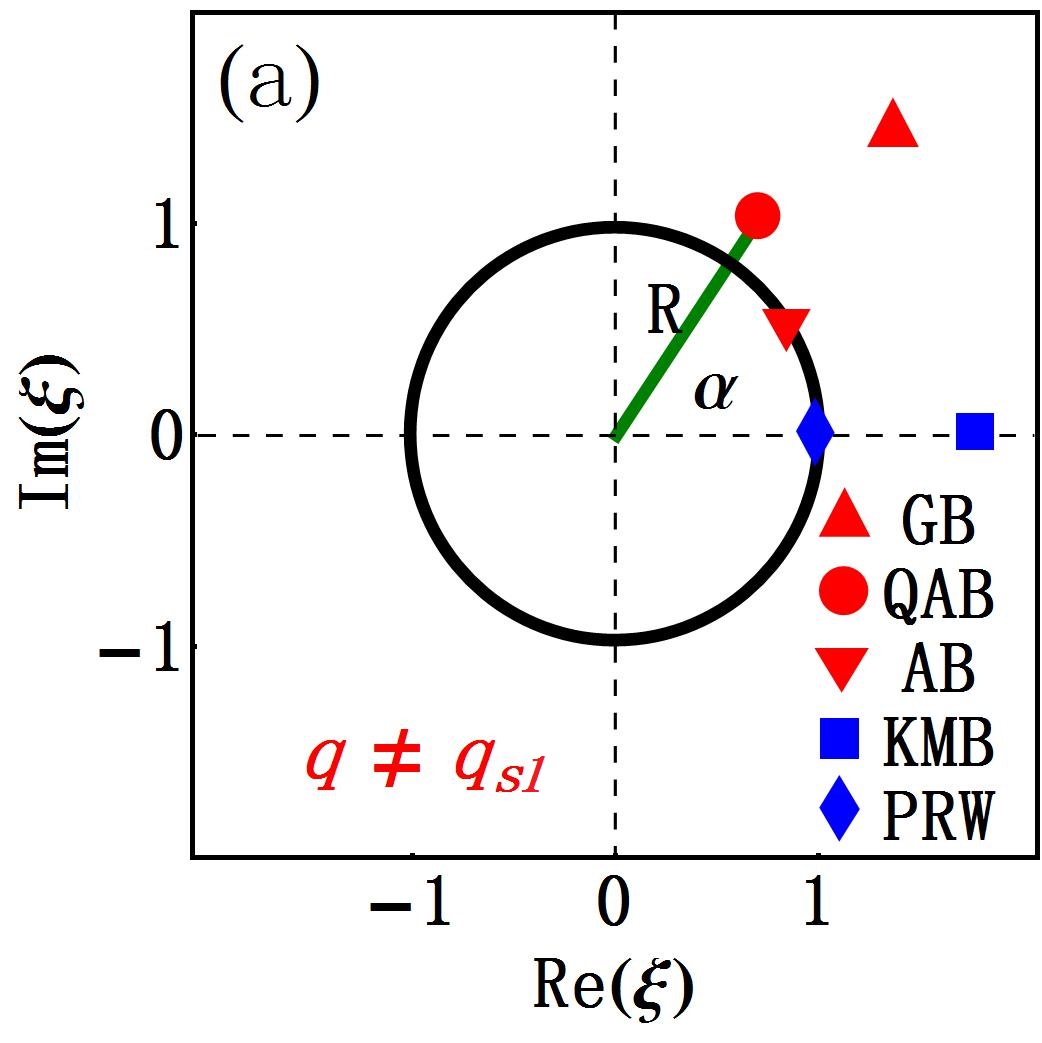}}
\subfigure{\includegraphics[height=42mm,width=42mm]{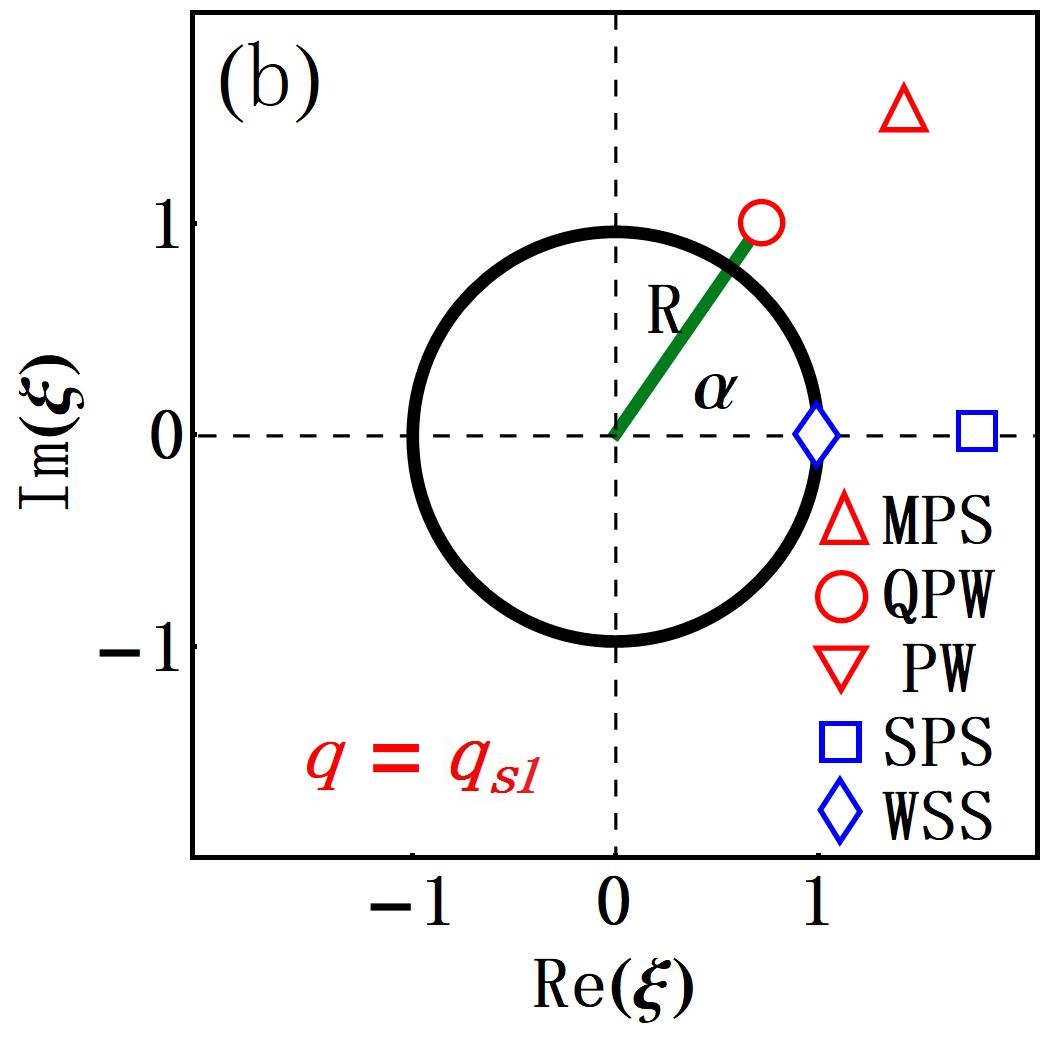}}
\caption{(color online) Phase diagrams of fundamental nonlinear modes in the $Re(\xi)$-$Im(\xi)$ plane ($\xi=Re^{i\alpha}$) with (a) $q\neq q_{s1}$ and (b) $q=q_{s1}$. (a) shows well-known breathing modes including ``GB'' (general breather), ``QAB'' (quasi-Akhmediev breather), ``AB'' (Akhmediev breather), ``KMB'' (Kuznetsov-Ma breather), and ``PRW'' (Peregrine rogue wave). While for the same pole (i.e., the same $R$, $\alpha$), (b) displays non-breathing modes including ``MPS'' (multipeak soliton), ``QPW'' (quasi-periodic wave), ``PW'' (periodic wave), ``SPS'' (single-peak soliton), and ``WSS'' (W-shaped soliton). 
}\label{fig1}
\end{figure}

However, not all of these parameters are essential. The parameter $a$ merely rescales the wave amplitude and velocity. Thus, without losing generality, we set $a=1$.
Additionally, the phase parameters $\theta_1$, $\mu_1$ can be dropped from
the discussion for the types of fundamental excitations shown in Fig. \ref{fig1}.
But special attention is needed to cover the higher-order solution
where the multiphase parameters [$\theta_j$, $\mu_j$ ($j\geq2$)] play a key role in the collision structures of multi waves (see the next section).

The remaining three independent parameters $R$, $\alpha$, $q$ play a pivotal role in the properties of fundamental nonlinear modes.
Specifically, the general solution describes nonlinear modes with transversal size $\Delta t\sim1/(2\eta_r)=\left[a(R-1/R)\cos{\alpha}\right]^{-1}$, propagating on top of the continuous wave with group velocity $V_{gr}$ and phase velocity $V_{ph}$. The longitudinal and transversal oscillation periods are $D_z=\pi/(\eta_iV_{ph})$, $D_t=\pi/\eta_i$, respectively.




\begin{figure}[htb]
\centering
\subfigure{\includegraphics[height=22mm,width=28mm]{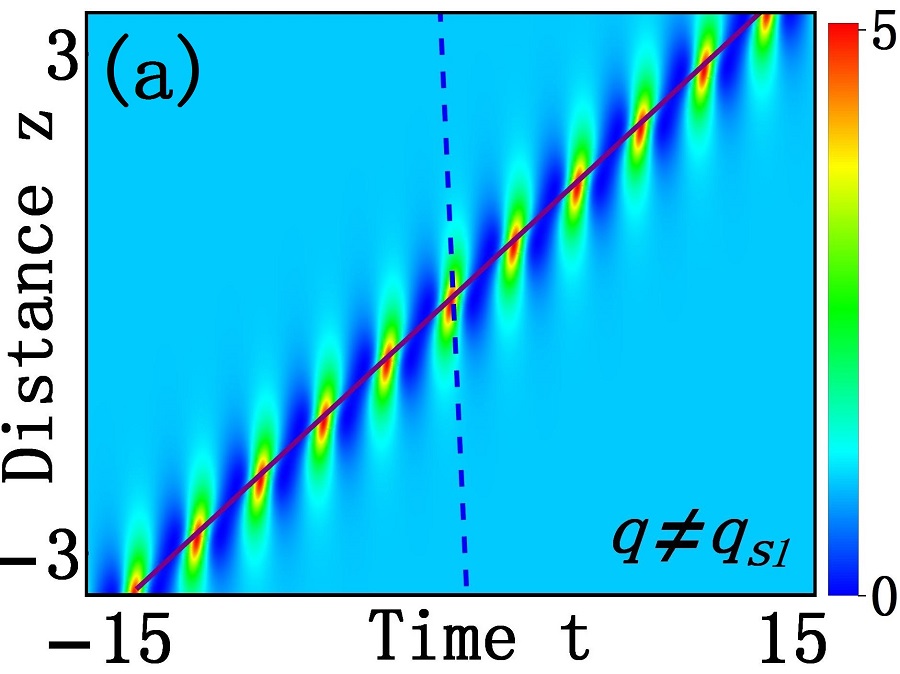}}
\subfigure{\includegraphics[height=22mm,width=28mm]{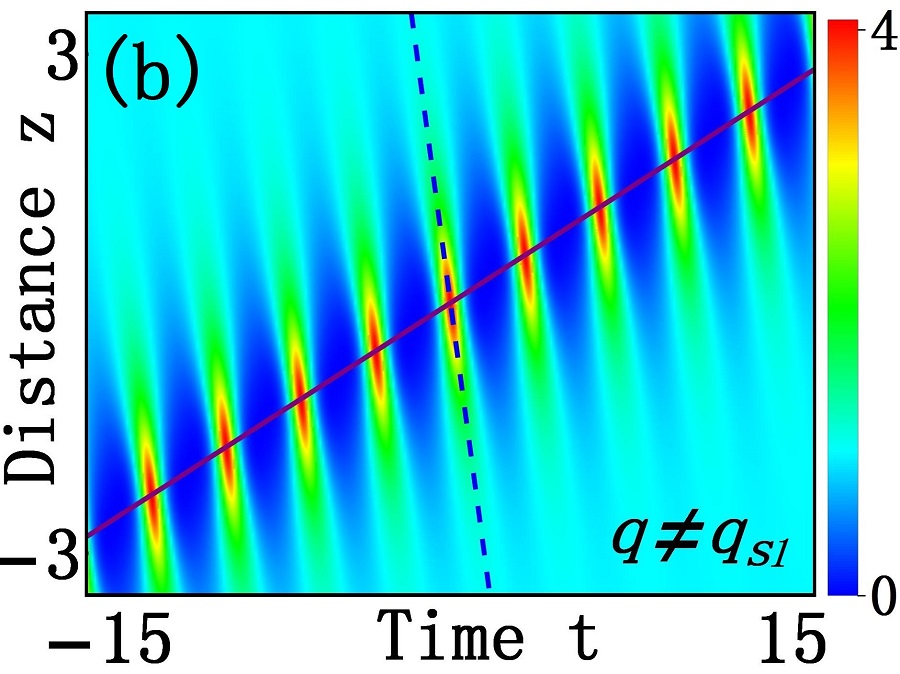}}
\subfigure{\includegraphics[height=22mm,width=28mm]{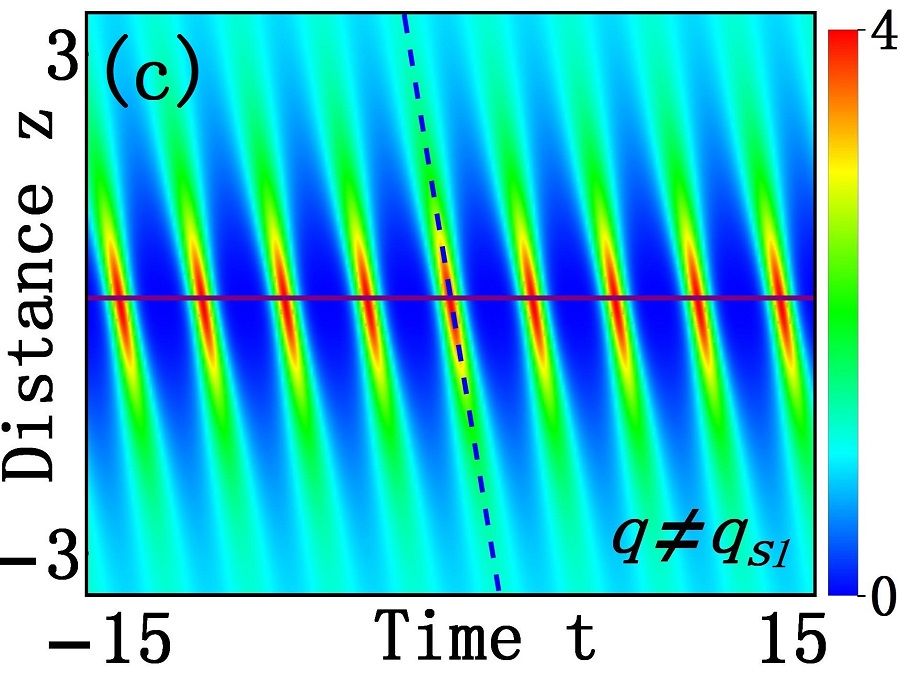}}
\subfigure{\includegraphics[height=22mm,width=28mm]{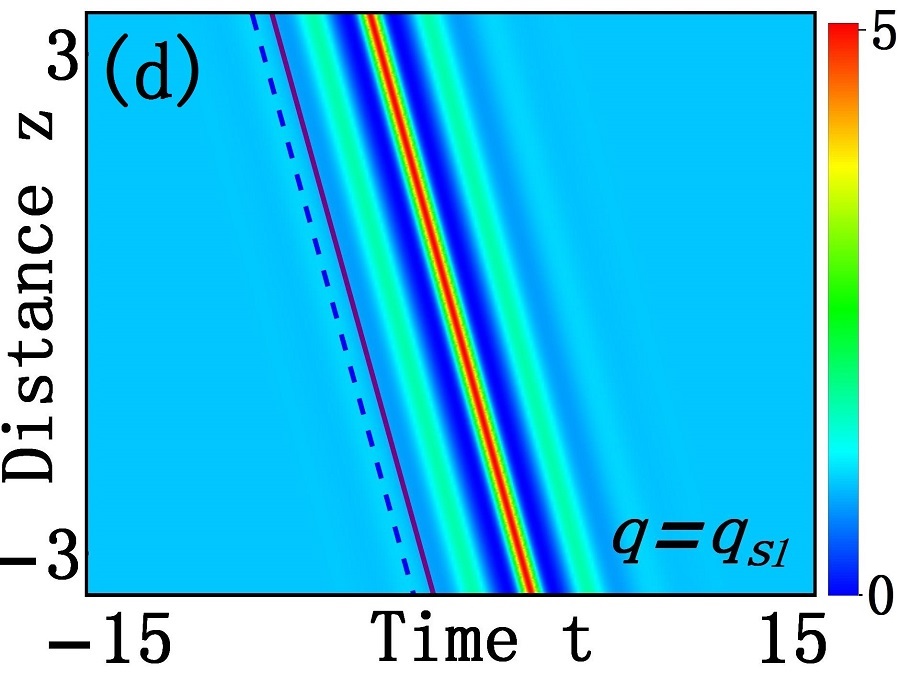}}
\subfigure{\includegraphics[height=22mm,width=28mm]{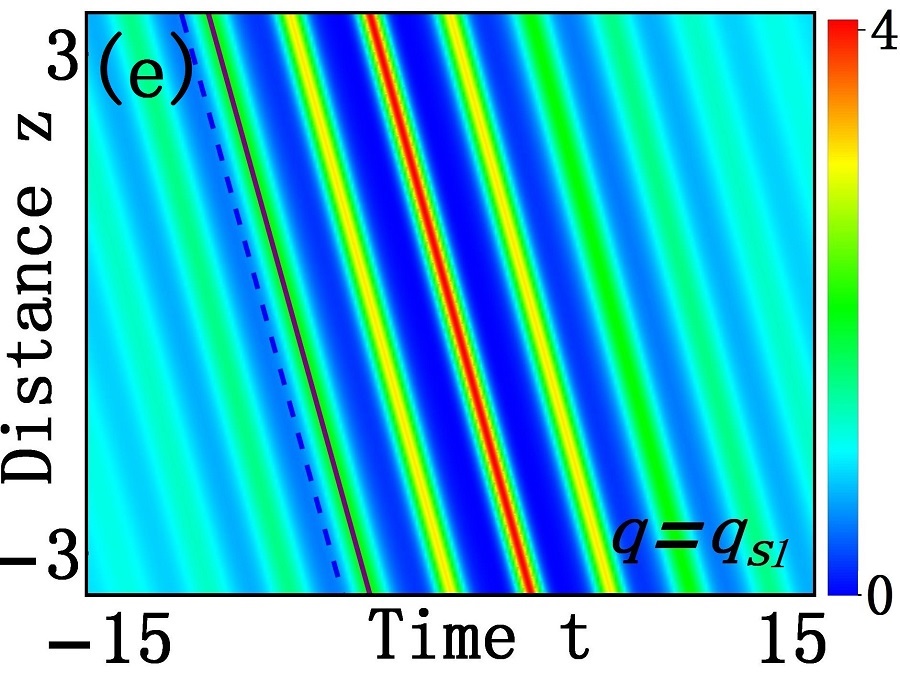}}
\subfigure{\includegraphics[height=22mm,width=28mm]{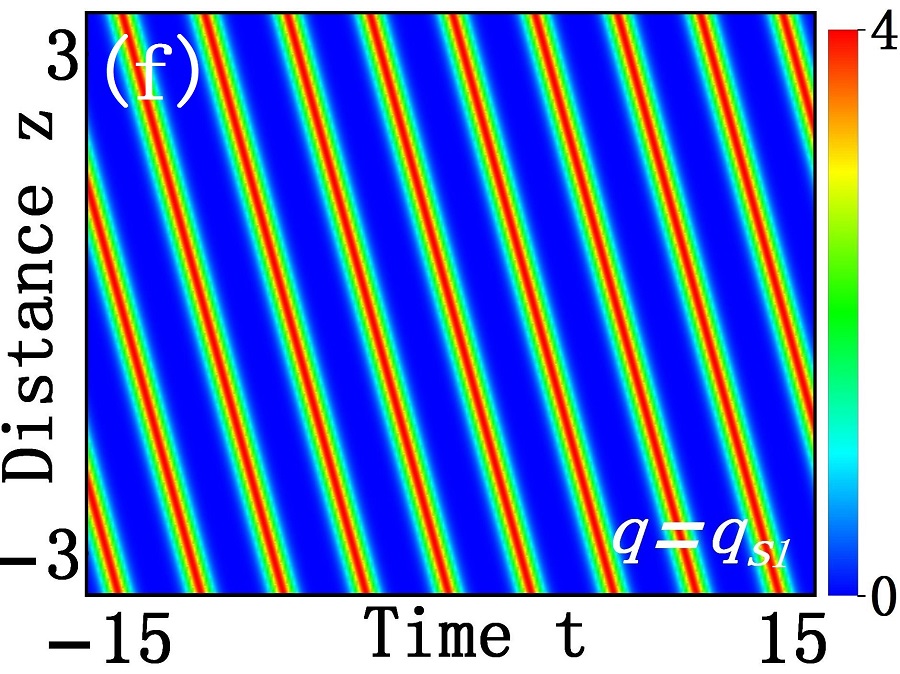}}
\caption{(color online) Intensity distribution $|u|^2$ of fundamental modes in $q\neq q_{s1}$ and $q=q_{s1}$ regimes as $R\rightarrow1$, given by Eq. (\ref{equs}).
Top row shows, in $q\neq q_{s1}$ (f.i., $q=0$) regime, (a) general breather ($R=2$), (b) quasi-Akhmediev breather ($R=1.2$), (c) Akhmediev breather ($R=1$).
Bottom row displays, in $q=q_{s1}$ regime, (d) multipeak soliton, (e) quasi-periodic wave, (f) periodic wave. Other parameters are $\alpha=\pi/3$, $\theta_1=\mu_1=0$, $a=1$.
The solid lines represent the group velocity $V_{gr}$, while the dashed lines describe the phase velocity $V_{ph}$.
}\label{fig2}
\end{figure}

An interesting finding is that the general nonlinear excitation exhibits significantly different features
as the change of $q$ leads to $V_{gr}\neq V_{ph}$ and $V_{gr}=V_{ph}$. Thus, a \textit{critical frequency}, i.e., $q_{s1}=-\frac{1}{6\beta}(1+8\varrho_1\beta)$, which is induced by the higher-order effects, is extracted.
We shed light on the complexity of fundamental modes via phase diagrams in the $Re(\xi)$-$Im(\xi)$ plane with $q\neq q_{s1}$ and $q=q_{s1}$ in Fig. \ref{fig1}.
Remarkably, if $q\neq q_{s1}$, the solution describes nonlinear modes displaying breathing property on the background $u_0$, which are known as breathers [see Fig. \ref{fig2}(a-c)].
Breathers in this particular case are consistent with the results in the standard NLSE \cite{SR1,SR2,SR3}.
However, instead if $q=q_{s1}$, nonlinear modes exhibit non-breathing profile along the propagation direction, which implies that the breathing property is gone completely [see Fig. \ref{fig2}(d-f)].
Fundamental modes in this interesting case appear as a result of the higher-order effects, which are of special importance.
Note also that this phase diagram is more general than that in the complex modified Korteweg-de Vries system \cite{HSR1}, since the choice of phase parameters ($\theta_1$, $\mu_1$) leads to the non-breathing waves with different structures. Indeed, one can readily confirm that the multipeak solitons exhibit structure symmetry and asymmetry when $\theta_1, \mu_1=0$ and $\theta_1, \mu_1\neq0$, respectively. Moreover, the single-peak soliton contains the W-shaped soliton ($\theta_1, \mu_1=0$) and antidark soliton ($\theta_1, \mu_1\neq0$). We point out that this general phase diagram for fundamental nonlinear waves on a nonvanishing background is also valid for the fourth-order NLSE \cite{HSR2}.

\section{Characteristics of femtosecond SR modes}
\begin{figure}[htb]
\centering
\includegraphics[height=60mm,width=85mm]{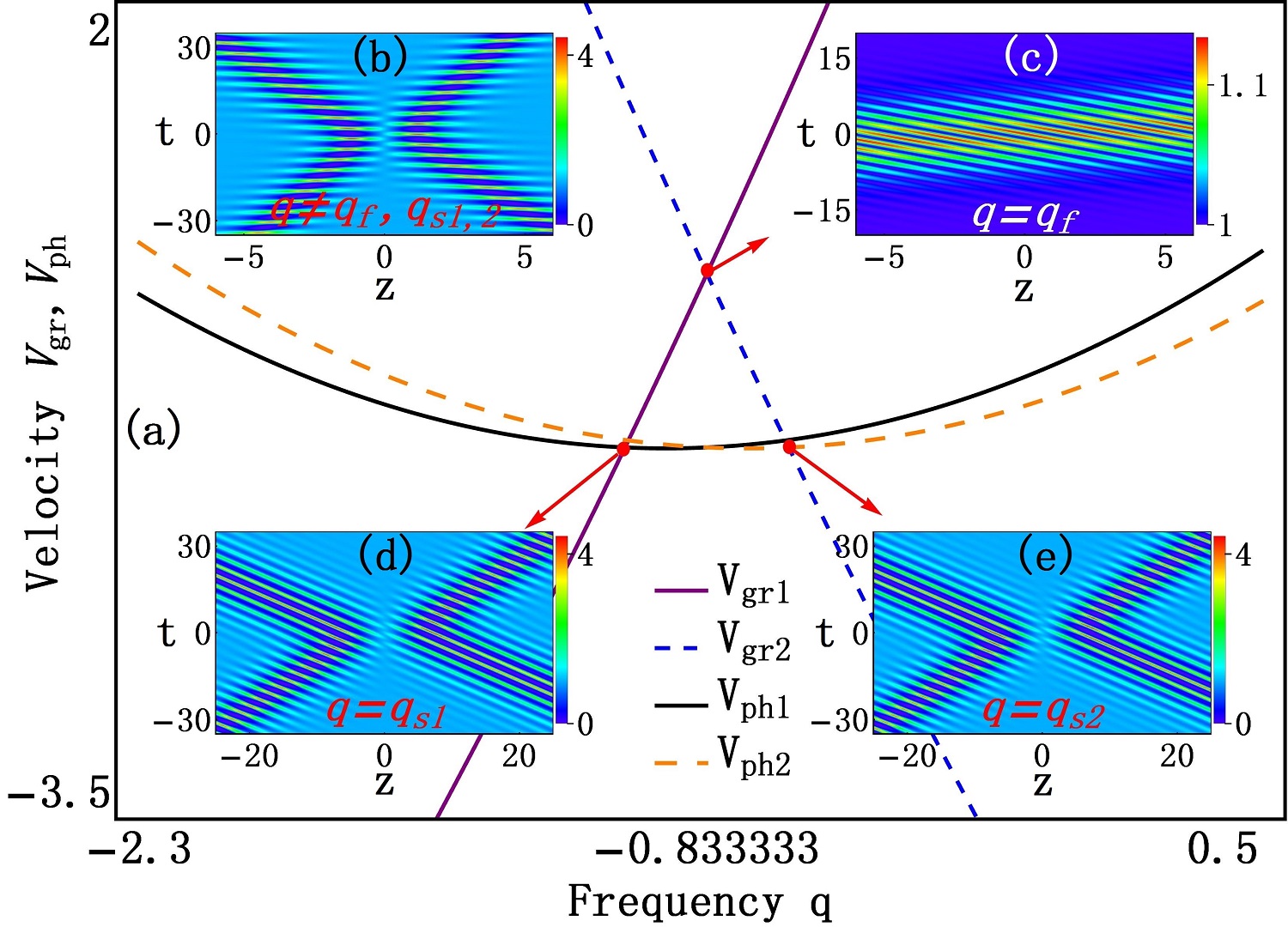}
\caption{(color online) (a) Evolution of $V_{grj}$, $V_{phj}$ $(j=1,2)$ with $q$.
The intersections at $q=q_f, q_{sj}$, where $q_f=-\frac{1}{6\beta}$, $q_{sj}=-\frac{1}{6\beta}(1+8\varrho_{j}\beta)$ reflect the nontrivial properties of SR modes, which are absent in the standard NLSE.
The insets, from bottom to top, show (b) the standard SR breather when $V_{gr1}\neq V_{gr2}$, $V_{phj}\neq V_{grj}$ (thus $q\neq q_f, q_{sj}$, we choose $q=2q_{s1}$), (c) the full-suppression SR mode when $V_{gr1}=V_{gr2}$ (thus $q=q_f$), (d) the half-transition SR mode when $V_{gr1}=V_{ph1}$, $V_{gr2}\neq V_{ph2}$ (thus $q=q_{s1}$), (e) the half-transition SR mode when $V_{gr2}=V_{ph2}$, $V_{gr1}\neq V_{ph1}$, (thus $q=q_{s2}$). Other parameters are $a=1$, $R=1.2$, $\alpha=\pi/3$, and $\theta_1=\theta_2=\pi/2$.}\label{fig3}
\end{figure}

Nonlinear superposition between the fundamental modes reported above reveal nontrivial coexistence and interaction of various types of fundamental modes, which attracts much attention \cite{n7,n8,n9,n10}.
However, our interest is confined to a novel superposition case, i.e., SR mode.
For the details, let us consider the simplest but very important case of SR one-pair solution (i.e., two-wave solution) with $R_1=R_2=R=1+\varepsilon$ ($\varepsilon\ll1$), $\alpha_1=-\alpha_2=\alpha$. The corresponding exact expressions can be obtained via the iteration of Darboux transformation which is given by
\begin{eqnarray}
u=u_0\left[1-4\rho \varrho_1\frac{(i\varrho_1-\rho)\Xi_1
+(i\varrho_1+\rho)\Xi_2}
{a(\rho^2\Xi_3+\varrho_1^2\Xi_4)}\right],\label{equsr}
\end{eqnarray}
where
\begin{eqnarray}
\Xi_1&=&\varphi_{21}\phi_{11}+\varphi_{22}\phi_{21},~\Xi_2=\varphi_{11}\phi_{21}+\varphi_{21}\phi_{22},\nonumber\\
\Xi_3&=&\varphi_{11}\phi_{22}-\varphi_{21}\phi_{12}-\varphi_{12}\phi_{21}+\varphi_{22}\phi_{11},\nonumber\\
\Xi_4&=&(\varphi_{11}+\varphi_{22})(\phi_{11}+\phi_{22}),\nonumber
\end{eqnarray}
and
\begin{eqnarray}
\phi_{jj}&=&\cosh(\Theta_2\mp i\psi)-\cos(\Phi_2\mp\alpha),\nonumber\\
\varphi_{jj}&=&\cosh(\Theta_1\mp i\psi)-\cos(\Phi_1\pm\alpha),\nonumber\\
\phi_{j3-j}&=&\pm i\cosh(\Theta_2\mp i\alpha)\mp i\cos(\Phi_2\mp \psi),\nonumber\\
\varphi_{j3-j}&=&\pm i\cosh(\Theta_1\pm i\alpha)\mp i\cos(\Phi_1\mp\psi),\nonumber
\end{eqnarray}
\begin{gather}
\Theta_j=2\eta_{rj}(t-V_{grj}z)+\mu_j,~\Phi_j=2\eta_{ij}(t-V_{phj}z)-\theta_j, \nonumber \\
V_{grj}=v_{1j}+v_{2j}\frac{\eta_{ij}}{\eta_{rj}}, ~~V_{phj}=v_{1j}-v_{2j}\frac{\eta_{rj}}{\eta_{ij}},\nonumber \\
v_{1j}=q+\varrho_j-\beta(2a^2-3q^2+4\rho^2-6q\varrho_j-4\varrho_j^2),\nonumber\\
v_{2j}=-6\beta q_{sj}\rho(1-q/q_{sj}),~q_{sj}=-\frac{1}{6\beta}(1+8\varrho_j\beta),\nonumber\\
\varrho_2=-\varrho_1,~\eta_{r1}=\eta_{r2}=\eta_{r},~\eta_{i1}=-\eta_{i2}=\eta_{i}.\nonumber
\end{gather}
where $\theta_{j}$, $\mu_{j}$ $(j=1,2)$ are phase shift parameters determining the collision property of two waves.
Specifically, both $\theta_{j}$ and $\mu_{j}$ impact the wave property at the collision location, but $\mu_{j}$ also determine the
collision location in the $z$-$t$ plane. Thus, without losing generality, we can set $\mu_{1,2}=0$, implying the collision emerges at $(z,t)=(0,0)$.
In this case, the remaining parameters $\theta_{j}$ determine the degree of complexity of
the wave profile around $(z,t)=(0,0)$. In particular, when $\theta_{1}+\theta_{2}=\pi$, the classical higher-order rogue waves \cite{BC} emerging around $(z,t)=(0,0)$ is gone, instead a novel quasiannihilation phenomenon of two quasi-Akhmediev modes, i.e., the standard SR modes, is observed [see, f.i., Fig. \ref{fig3}(a)].
Note that this standard SR mode consists of two quasi-Akhmediev waves with $V_{gr1}\neq V_{gr2}$.

Once the parameters $\varepsilon$, $\mu_{j}$ $(=0)$, $\theta_{j}$ (f.i., $\theta_{j}=\pi/2$) are fixed, we are left only with the free parameter $q$.
The latter impacts both $V_{grj}$ and $V_{phj}$ which will play a key rule in the characteristics of SR modes (see Fig. \ref{fig3}).
Indeed, as reported above, a quasi-Akhmediev breather can be converted to a quasi-periodic wave when $V_{gr}=V_{ph}$ (thus $q=q_{s1}$).
On the other hand, one should note that the standard SR mode consists of two quasi-Akhmediev breathers with $V_{gr1}\neq V_{gr2}$.
Thus, in what follows we shall reveal the characteristics of SR modes of Eq. (\ref{equ1}) by the properties of $V_{grj}$, $V_{phj}$ with $q$ in Eq. (\ref{equsr}).

Figure \ref{fig3} shows the evolution characteristics of $V_{grj}$, $V_{phj}$ with $q$. We highlight the features of SR modes in the insets. In particular, the intersections at $q=q_f, q_{sj}$ reflect the unique properties of SR modes, which are absent in the standard NLSE.

The standard SR breather [see the inset, Fig. \ref{fig3}(b)] exists when $V_{gr1}\neq V_{gr2}$, $V_{grj}\neq V_{phj}$ which implies $q\neq q_f, q_{sj}$, where $q_f=-\frac{1}{6\beta}$, $q_{sj}=-\frac{1}{6\beta}(1+8\varrho_{j}\beta)$.
In this case the expression (\ref{equsr}) yields the trivial generalization of the SR breather in the femtosecond regime.
As shown, the classical SR mode describes the quasiannihilation of breather collision at the line $z=0$ and the amplification of small localized perturbation [i.e., $u(0,t)$] as $z$ increases.

Instead, when $V_{gr1}=V_{gr2}$ (thus $q=q_f$), implying that two quasi-Akhmediev modes with the same group velocity in the $z$-$t$ plane, the
SR mode becomes a novel breather-complex state with small amplitude propagating along $z$ [Fig. \ref{fig3}(c)]. By comparison with the standard SR breather describing the amplification of small localized perturbation as $z$ increases [Fig. \ref{fig3}(b)],
we term it \textit{full-suppression} SR mode. One should note that, in general, a small perturbation on a continuous wave background displays the feature of MI, namely, it will be amplified and distorted. Surprisingly,
the full-suppression mode describes a crucial \textit{non-amplifying} nonlinear dynamics of localized small perturbations on a continuous wave background along $z$. Namely, the small perturbation \textit{cannot} be amplified exponentially, but propagates along $z$ with small oscillations. It can be inferred that this phenomenon stems from the regime
where the MI is suppressed completely when $q=q_f$.

On the other hand, when $V_{grj}=V_{phj}$, $V_{gr3-j}\neq V_{ph3-j}$ (thus $q=q_{sj}$), we observe the half-transition phenomenon of SR modes. Namely,
one quasi-Akhmediev breather is converted to a quasi-periodic wave, while the other remains the nature of the breather [Figs. \ref{fig3}(d) and (e)].
Interestingly, the quasiannihilation phenomenon around $(z,t)=(0,0)$ always holds for the half-transition process. Namely, a quasi-Akhmediev breather and a quasi-periodic wave
collide and form a small-amplitude wave structure (i.e., quasiannihilation) around $(z,t)=(0,0)$; then this small-amplitude perturbation is amplified exponentially and eventually becomes a mix of quasi-Akhmediev and quasi-periodic waves. Note also that the quasi-periodic wave undergoes redistribution of power between the peaks before and after the quasiannihilation collision.
This stems from the phase shift arising from the collision. Note that the half-transition SR mode is a general SR structure induced by higher-order effects which can be observed in the other integrable systems \cite{HSR1,HSR2}. However, the full-suppression SR mode may have different structures depending on different higher-order effects \cite{HSR1,HSR2}.


\section{Different SR modes generated from an identical initial perturbation}
\begin{figure}[htb]
\centering
\subfigure{\includegraphics[height=21mm,width=42mm]{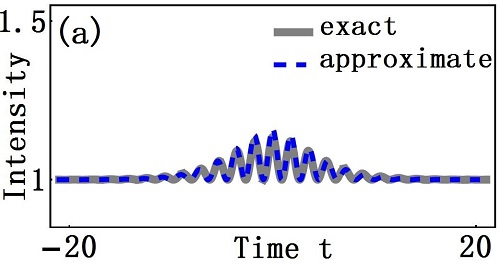}}
\subfigure{\includegraphics[height=21mm,width=42mm]{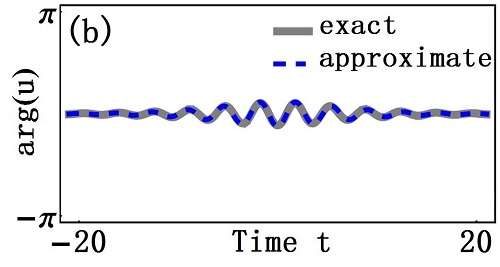}}
\subfigure{\includegraphics[height=33mm,width=21mm]{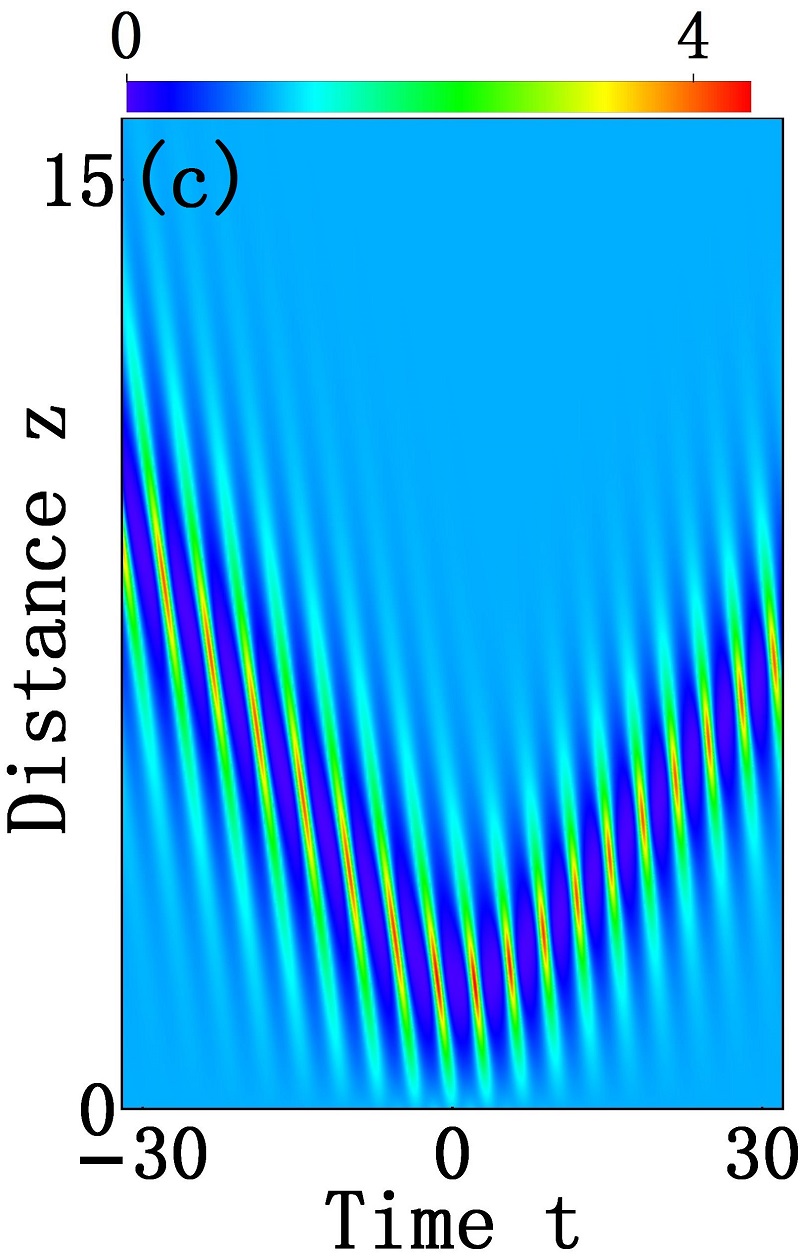}}
\subfigure{\includegraphics[height=33mm,width=21mm]{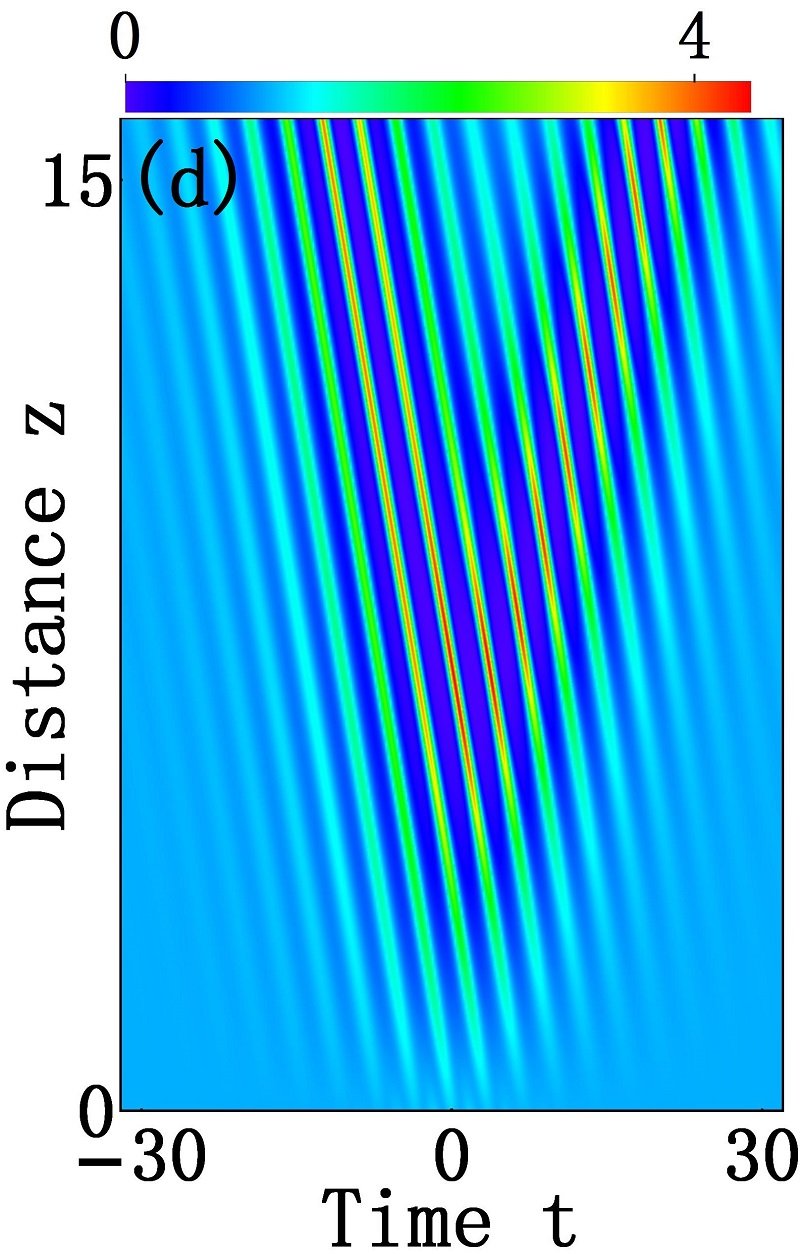}}
\subfigure{\includegraphics[height=33mm,width=21mm]{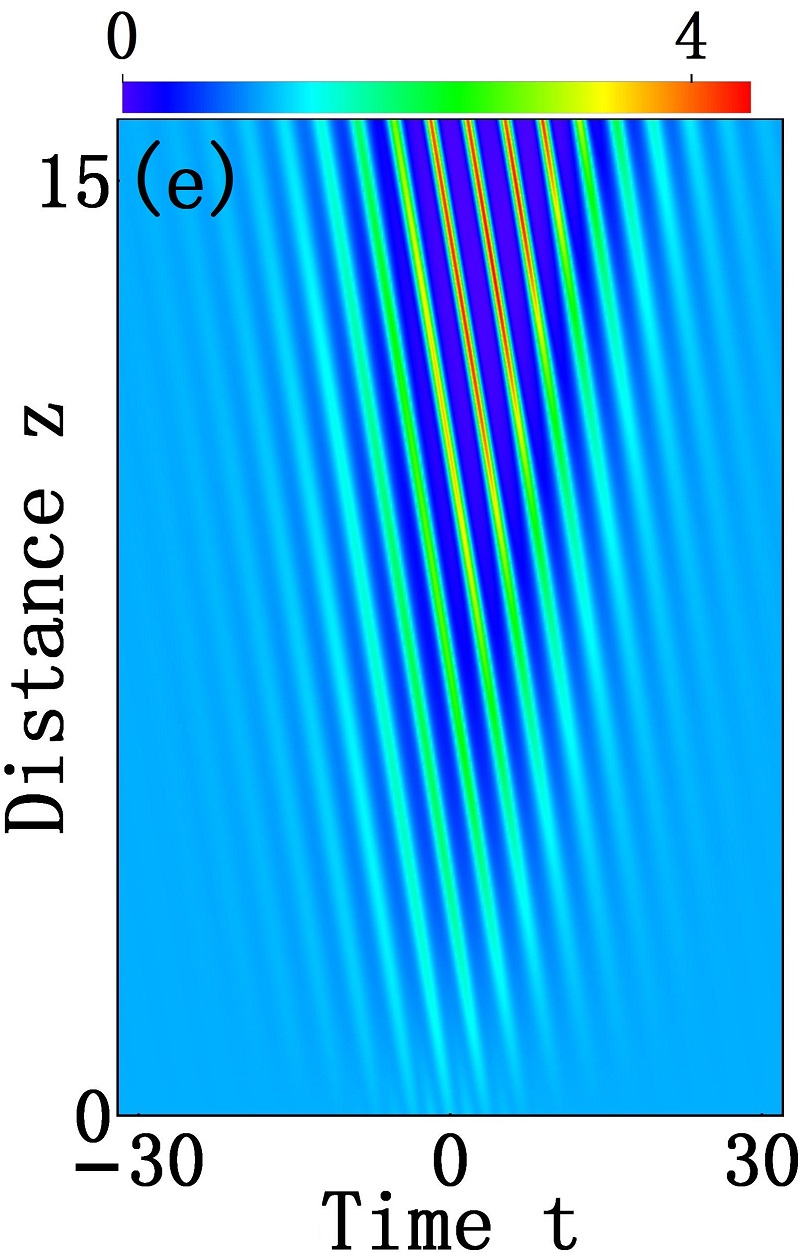}}
\subfigure{\includegraphics[height=33mm,width=20mm]{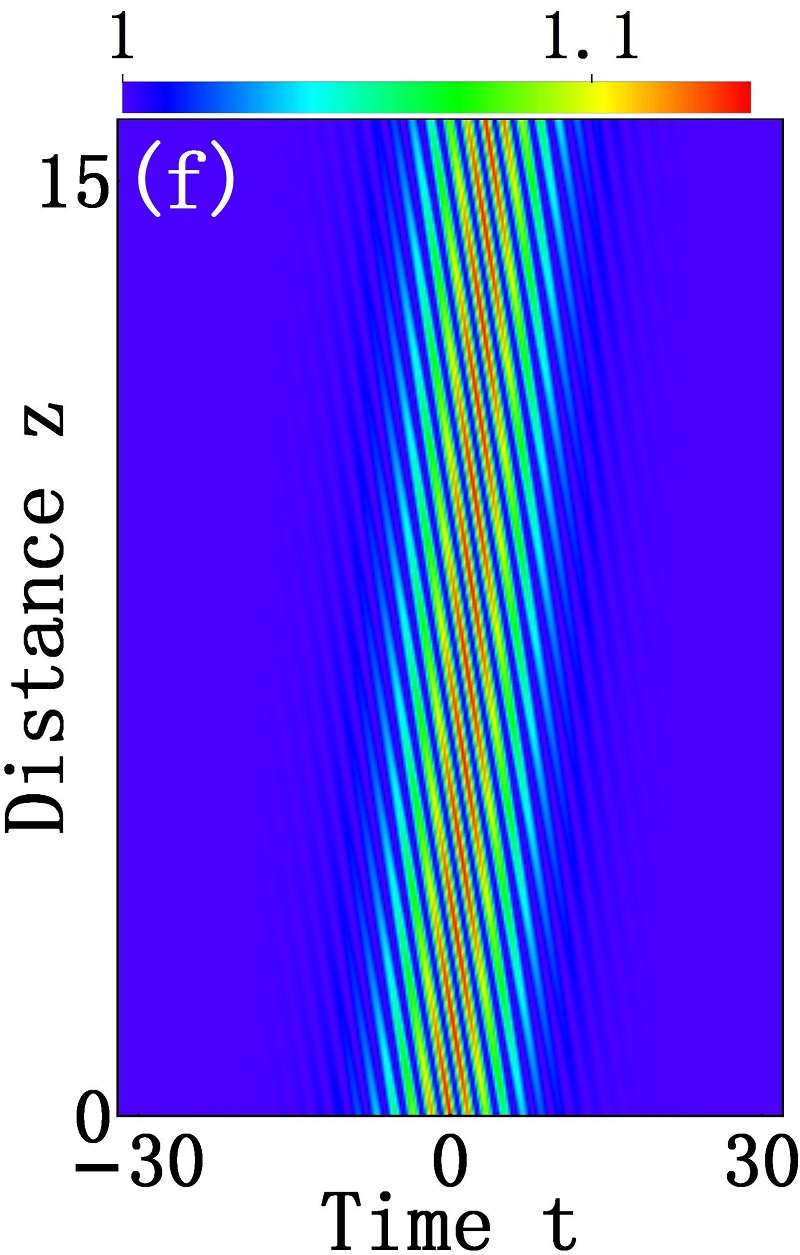}}
\caption{(color online) Top row: the intensity $|u|^2$ (a) and phase $arg(u)$ (b) profiles of the initial small-amplitude perturbation, extracting from the exact solution (\ref{equsr})
at $z=0$ and the approximate solution (\ref{eqp}), $\delta u$. The setup is the same as in Fig. \ref{fig3}. Bottom row: nonlinear evolution ($|u|^2$) of the identical initial perturbation with different values of $q$, given by Eq. (\ref{equsr}), (c) $q=1.5q_{s1}$, (d) $q=q_{s1}$, (e) $q=0.9q_{s1}$, (f) $q=q_{f}$.
}\label{fig4}
\end{figure}

The next step of interest and significance is to reveal the characteristic of nonlinear evolution of the initial small-amplitude perturbation $\delta u$, which is derived from the exact solution (\ref{equsr}) at $z=0$ [i.e., $u(0,t)=(a+\delta u)e^{i\theta}$],
Following the procedure described in Refs. \cite{SR1,SR2,SR3}, $\delta u$ can be written as
\begin{eqnarray}
&&\delta u\approx-\frac{4ia\varepsilon\cosh{(i\alpha)}\cos{\left(2at\sin{\alpha}\right)}}{\cosh{(2a\varepsilon t \cos{\alpha})}}.
\label{eqp}
\end{eqnarray}
The expression (\ref{eqp}) is an approximate formula to describe the localized small perturbation.
$\delta u$ is localized along $t$ and possesses a perturbed frequency $2a\sin{\alpha}$. The intensity $|\delta u|^2$ is proportional to $\varepsilon$.
Hence we should let $\varepsilon\ll1$ to hold $|\delta u|^2\ll a^2$. Figures \ref{fig4}(a) and \ref{fig4} (b) show the comparison between the approximated
perturbation $\delta u$ and the exact profile in Fig. \ref{fig3}. 

On the other hand, one should note that $\delta u$ depends on parameters $\alpha$, $\varepsilon$, but has no connection with $q$. Thus once $\alpha$, $\varepsilon$ are fixed, the initial profile $\delta u$ is given. However, the subsequent nonlinear evolution of
the small-amplitude perturbation exhibits diversity depending on the value of $q$. Namely, different SR states can be developed from an identical initial perturbation. The property reported in the scalar fiber systems here is different from the vector complementary waves in the vector three-wave mixing system \cite{CRW}.

Figures \ref{fig4}(c)-\ref{fig4}(f) illustrate the involved nonlinear characteristics evolving from the identical initial perturbation $\delta u$ in Figs. \ref{fig4}(a) and \ref{fig4}(b)
through which a full-suppression mode forms, when the parameter $q$ is varied.
Specifically, as $q$ decreases from $1.5q_{s1}$ to $q_f$, implying $|V_{gr1}-V_{gr2}|\rightarrow0$, the nonlinear state of the identical initial perturbation
manifests as one-pair quasi-Akhmediev [Figs. \ref{fig4}(c) and \ref{fig4}(e)], a mix of quasi-Akhmediev and quasi-periodic [Figs. \ref{fig4}(d)], and full-suppression [Figs. \ref{fig4}(f)] modes. We also note that as $q\rightarrow q_f$, the growth rate of the perturbation amplitude decreases gradually.

\section{numerical simulations}
Let us then confirm the stability of different SR modes via numerical simulations.
We perform direct numerical simulations of Eq. (\ref{equ1}) by the split-step Fourier method.
In our previous result, the robustness of SR breathers has been verified numerically from the ideal initial states in the special complex modified Korteweg-de Vries system \cite{HSR1}.
In the following, we will study the robustness of different femtosecond SR breathers in two different ways.
Namely, we first study nonlinear evolution from the initial condition given by the approximate solution (\ref{eqp}) in both integrable and nonintegrable cases. Then we will reveal the property of SR breathers from various nonideal initial excitations.

Note that significant progresses have been recently made on the robustness of rogue waves in nonintegrable systems \cite{NM1,NM2}.
It reveals that nonintegrable systems can also admit quasi-integrable regimes where rogue waves exist.
On the other hand, it is interesting to answer whether the dynamics of SR breathers can be created by nonideal initial excitations.
This important problem has been recently addressed in the standard NLSE system \cite{NM3}. It demonstrated that the standard SR breather turns out to be an universal nonlinear excitations by solving the Zakharov-Shabat eigenvalue problem associated with some nonideal (Sech and Gaussian) initial states. Moreover, the SR breather dynamics has been also observed numerically from a noise perturbation [see Chap.7 in Ref. \cite{MIr3}].

\subsection{integrable and nonintegrable cases}
\begin{figure}[htb]
\centering
\subfigure{\includegraphics[height=33mm,width=21mm]{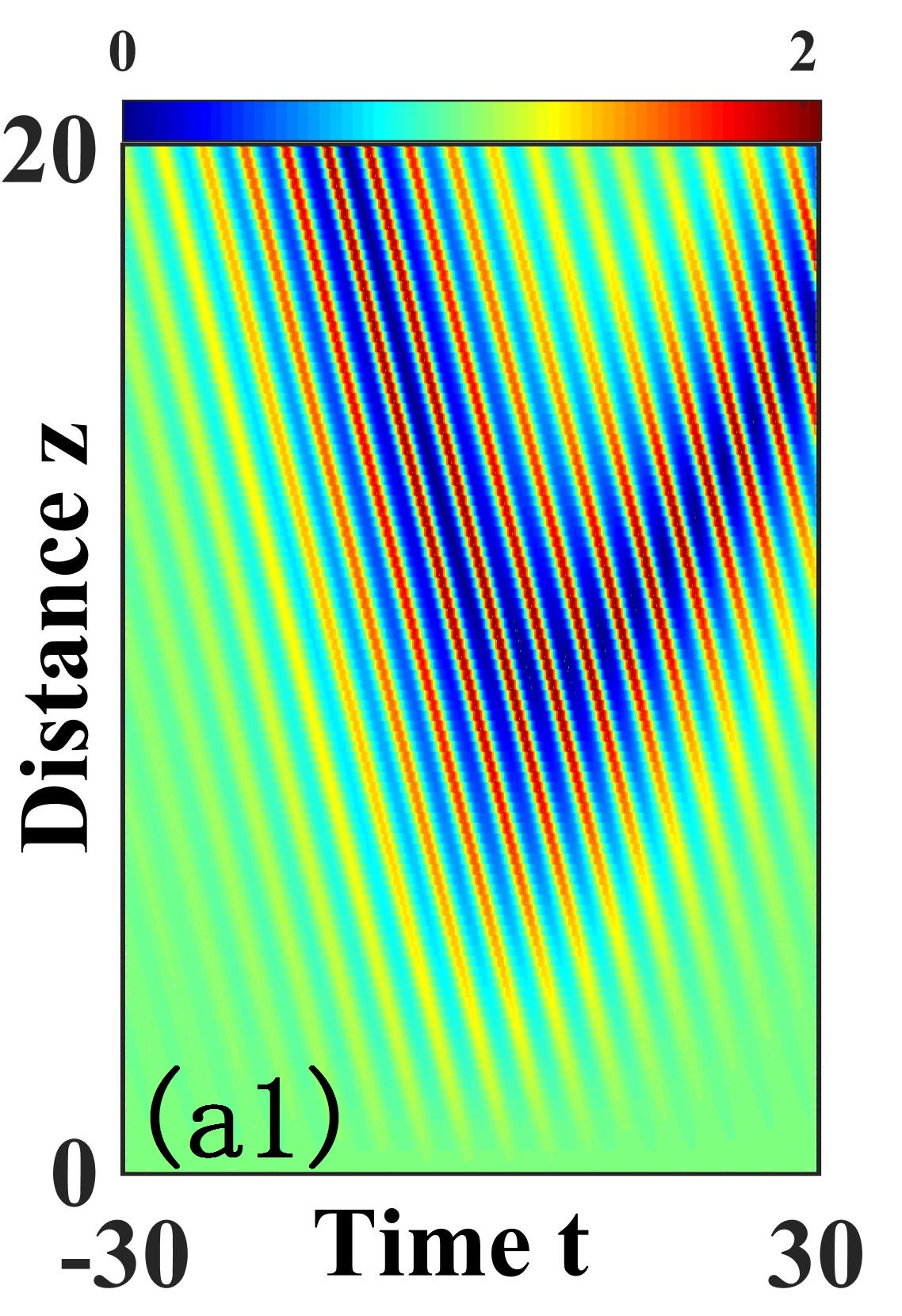}}
\subfigure{\includegraphics[height=33mm,width=21mm]{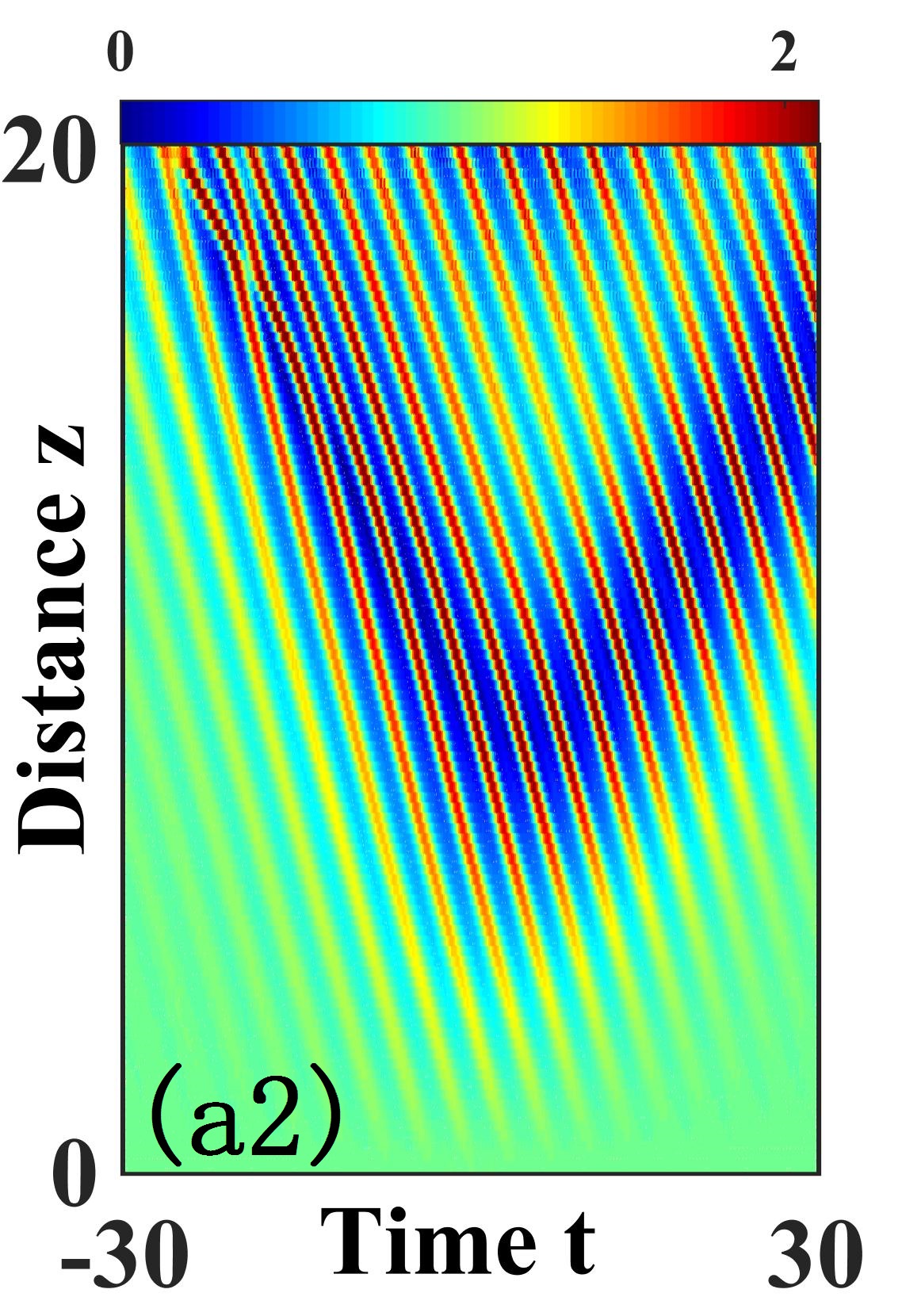}}
\subfigure{\includegraphics[height=33mm,width=21mm]{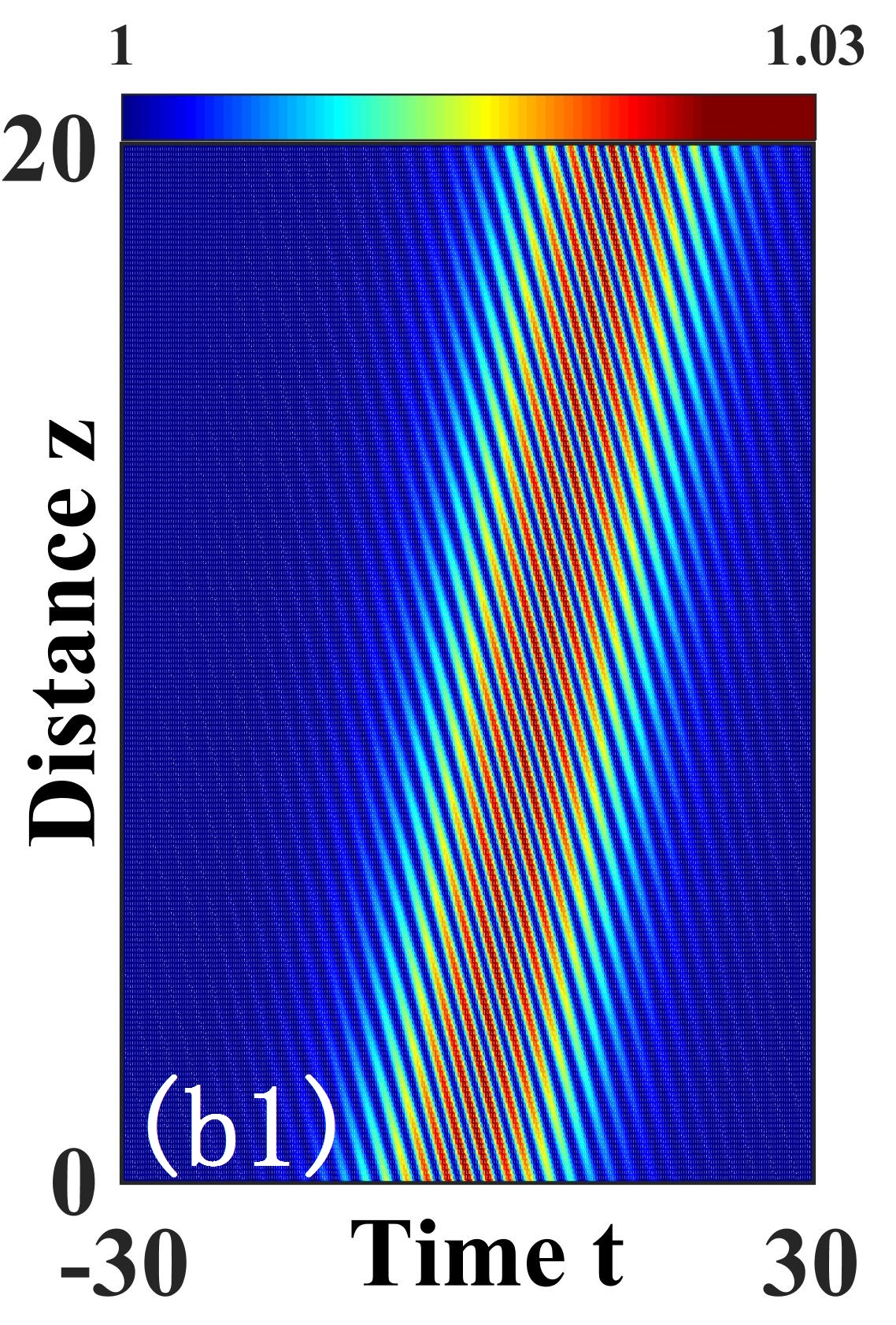}}
\subfigure{\includegraphics[height=33mm,width=20mm]{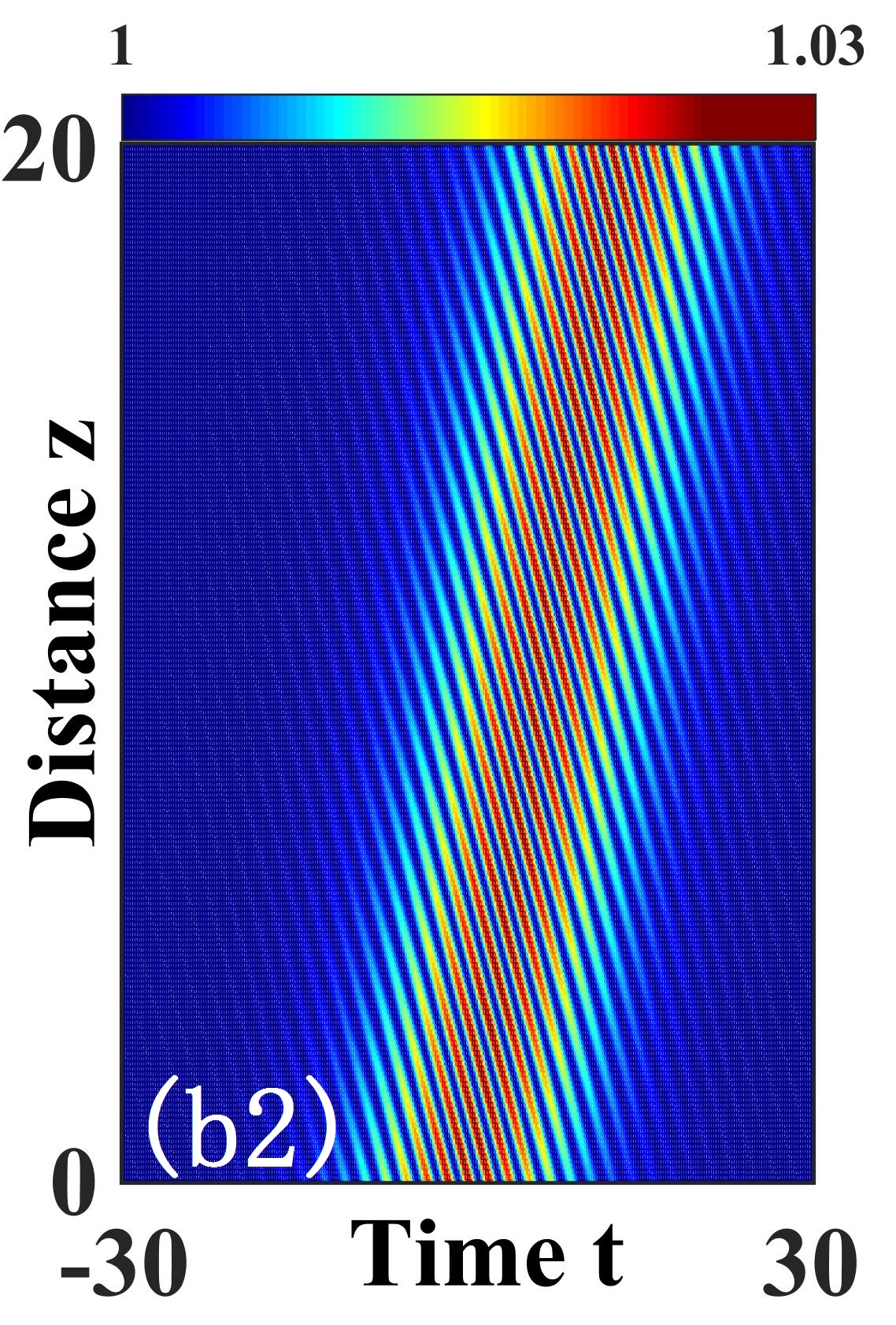}}
\caption{(color online) Numerical confirmation of (a) half-transition and (b) full-suppression SR modes in (1) integrable ($\Delta=0$) and (2) nonintegrable ($\Delta=0.1$) cases from the approximate solution (\ref{eqp}). The setup is the same as in Fig. \ref{fig4}, but $\beta=0.3$ and a smaller $\varepsilon=0.12$.
}\label{fig5}
\end{figure}

We consider the condition departing from the validity of the integrable case, i.e., $s+\gamma=0$, $s=(6+\Delta)\beta$, where $\Delta(\ll1)$ is a real parameter describing the deviation. The comparison of numerical results of the typical half-transition and full-suppression modes in integrable and nonintegrable cases (i.e., $\Delta=0$ and $\Delta\neq0$) is depicted in Fig. \ref{fig5}.

As shown in Figs. \ref{fig5}(a1) and \ref{fig5}(a2), the numerical results of half-transition SR modes are in good agreement with each other for the first fifteen propagation
distances. After that, the mode with $\Delta\neq0$ exhibits a breakup state [Fig. \ref{fig5}(a2)] arising from the deviations of the integrable condition, while the mode with $\Delta=0$ exhibits robustness over twenty propagation distances [Fig. \ref{fig5}(a1)]. In contrast, our numerical results of full-suppression SR modes indicate no collapse arising from the deviations of the integrable condition. Instead, stable propagation over twenty of propagation distances is observed [Figs. \ref{fig5}(b1) and \ref{fig5}(b2)]. This stems from that the full-suppression mode exists in a region with vanishing MI growth rate (see next section).

\subsection{nonideal initial excitations}

Next we confirm directly the robustness of nontrivial dynamics of femtosecond SR breathers induced by higher-order effects from various nonideal initial pulses.
To this end, the localized function in Eq. (\ref{eqp}) 
is replaced by a generalized localized form $f(t)$. Thus the nonideal initial perturbation is of the form
\begin{eqnarray}
&&\delta u=-if(t)\cos{\left(2at\sin{\alpha}\right)}.
\label{eqpn}
\end{eqnarray}
Here $f(t)$ denotes different types of localized functions which can be modulated to be close to the ideal state.
Eq. (\ref{eqpn}) is the simplest form of nonideal perturbations in Ref. \cite{NM3}.

\begin{figure}[htb]
\centering
\includegraphics[height=53mm,width=84mm]{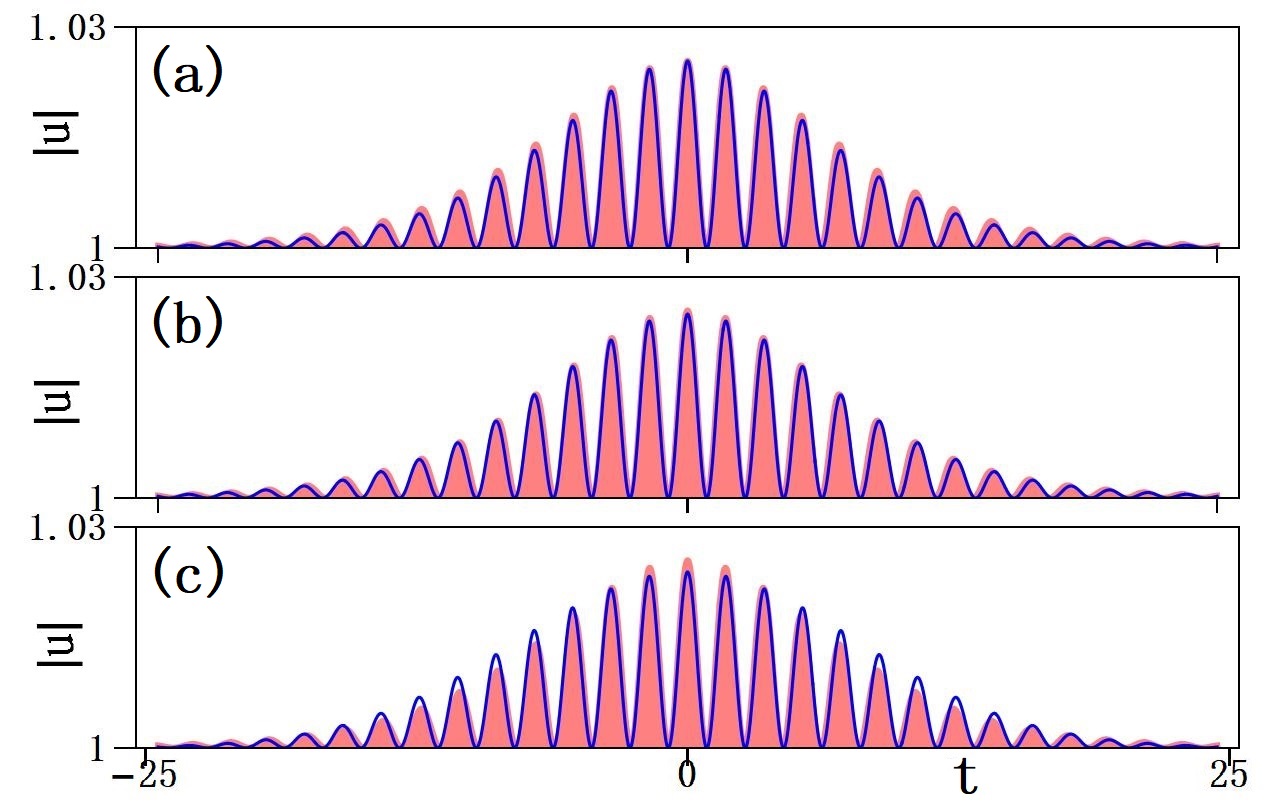}
\caption{(color online)  Amplitude profiles $|u(0,t)|$ (blue lines) of different nonideal small-amplitude initial states
 (a) the sech type $f(t)=0.227\textrm{sech}(0.118t)$,
(b) the Lorentzian type $f(t)=0.225/(1+0.003t^2)^2$, and (c) the Gaussian type $f(t)=0.22\exp(-t^2/256)$.
The pink regions represent the ideal initial excitation extracted from SR breathers.
The setup of the ideal initial state is the same as in in Fig. \ref{fig5}, but $\beta=0.2$.
}\label{figns}
\end{figure}
\begin{figure}[htb]
\centering
\subfigure{\includegraphics[height=31mm,width=23mm]{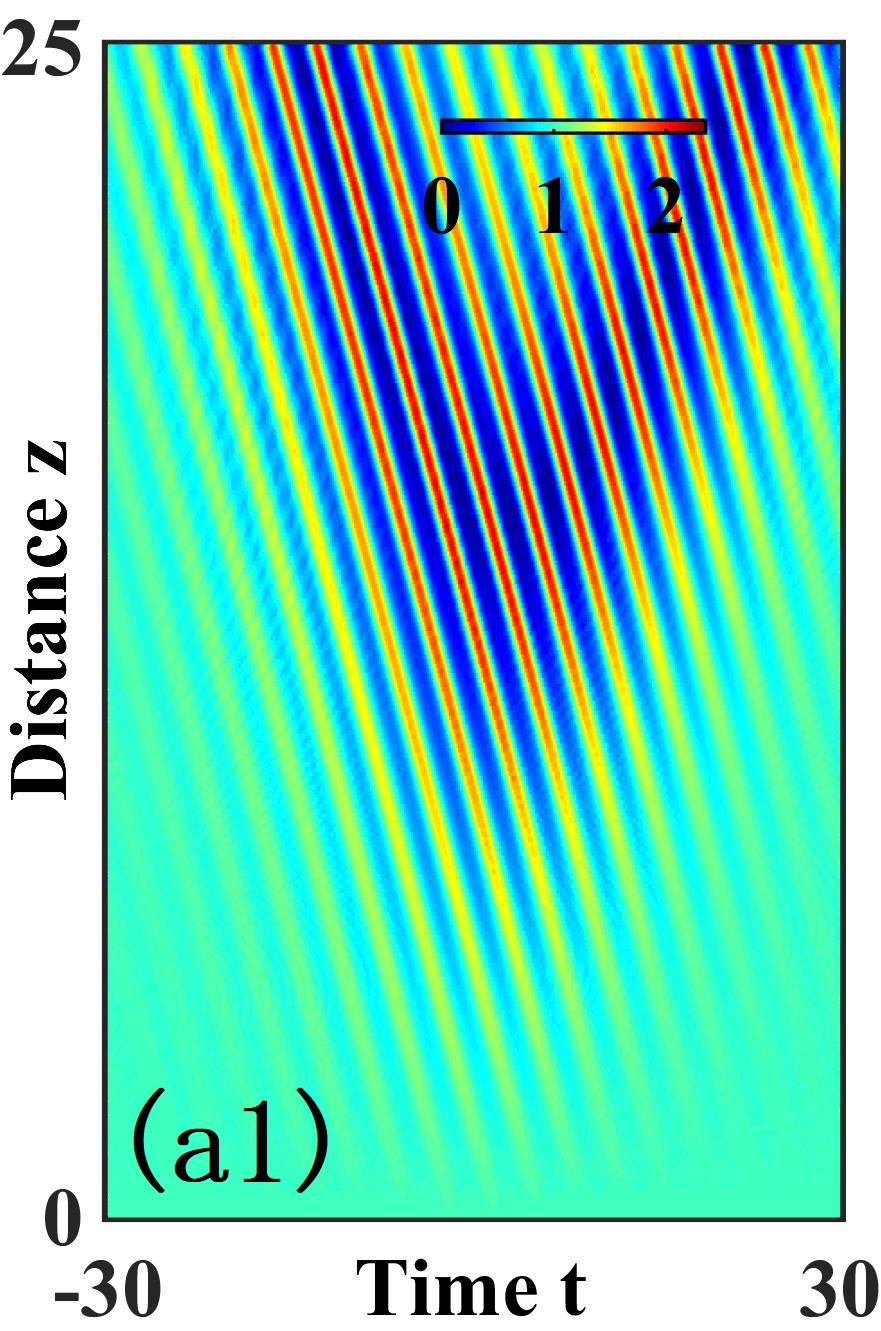}}
\subfigure{\includegraphics[height=31mm,width=23mm]{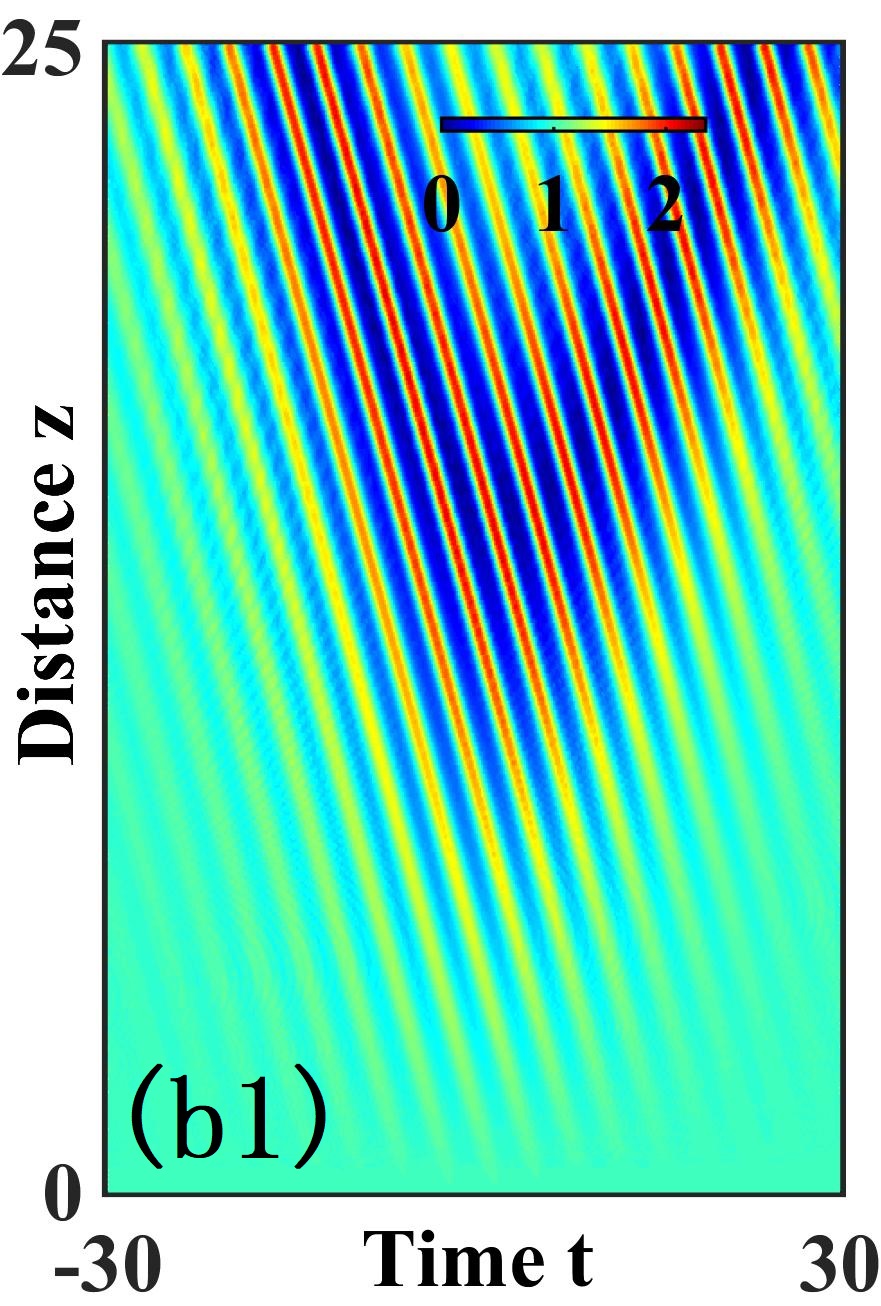}}
\subfigure{\includegraphics[height=31mm,width=23mm]{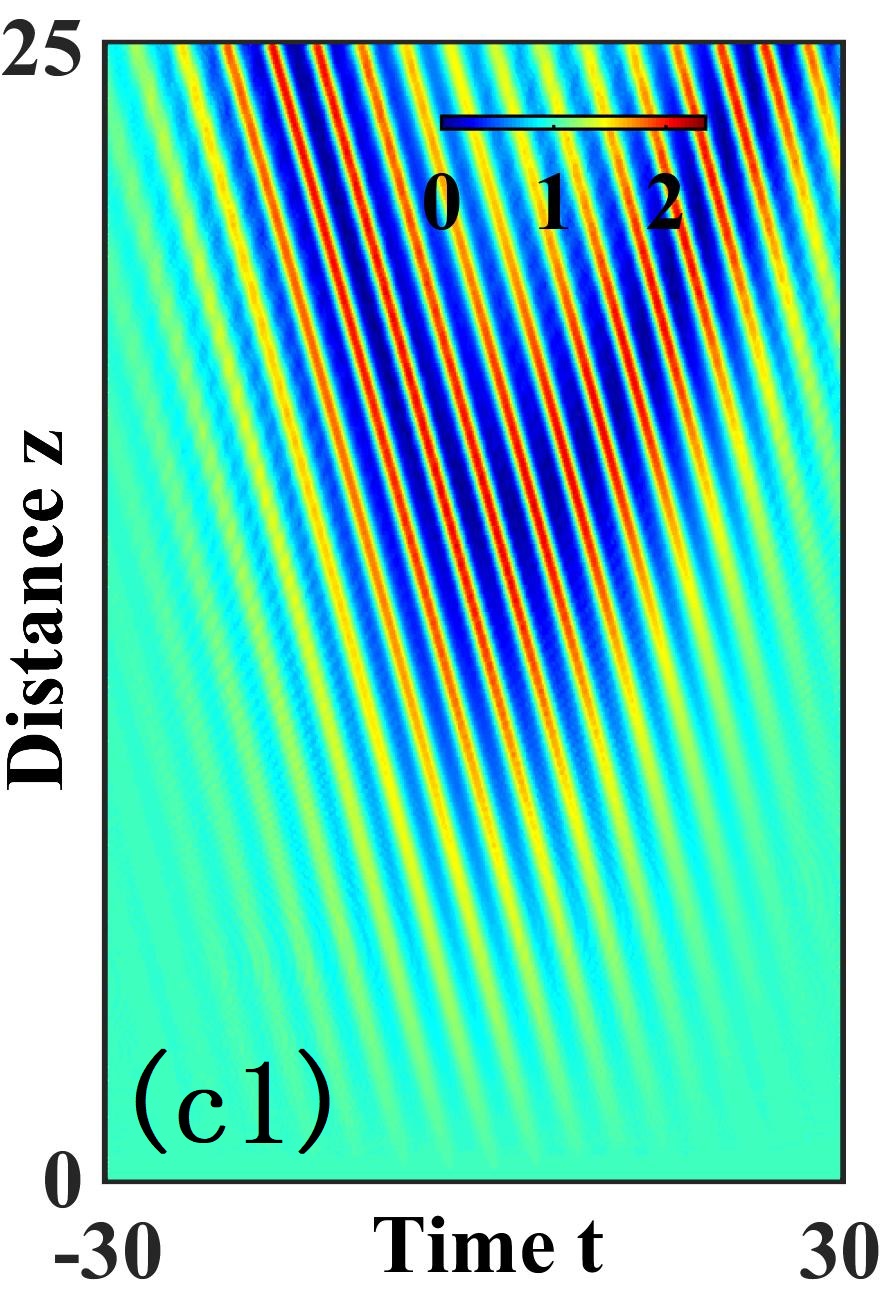}}
\subfigure{\includegraphics[height=31mm,width=23mm]{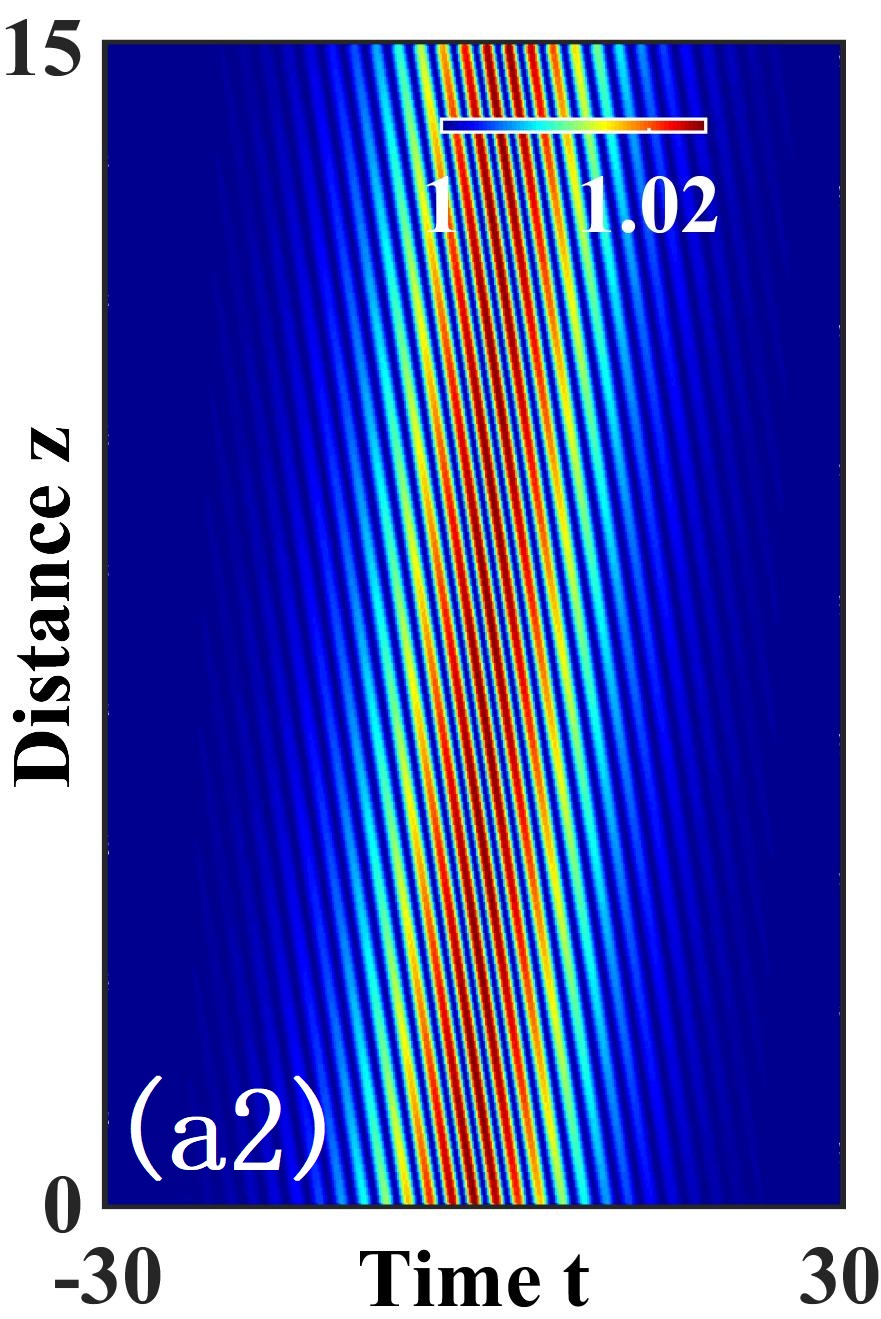}}
\subfigure{\includegraphics[height=31mm,width=23mm]{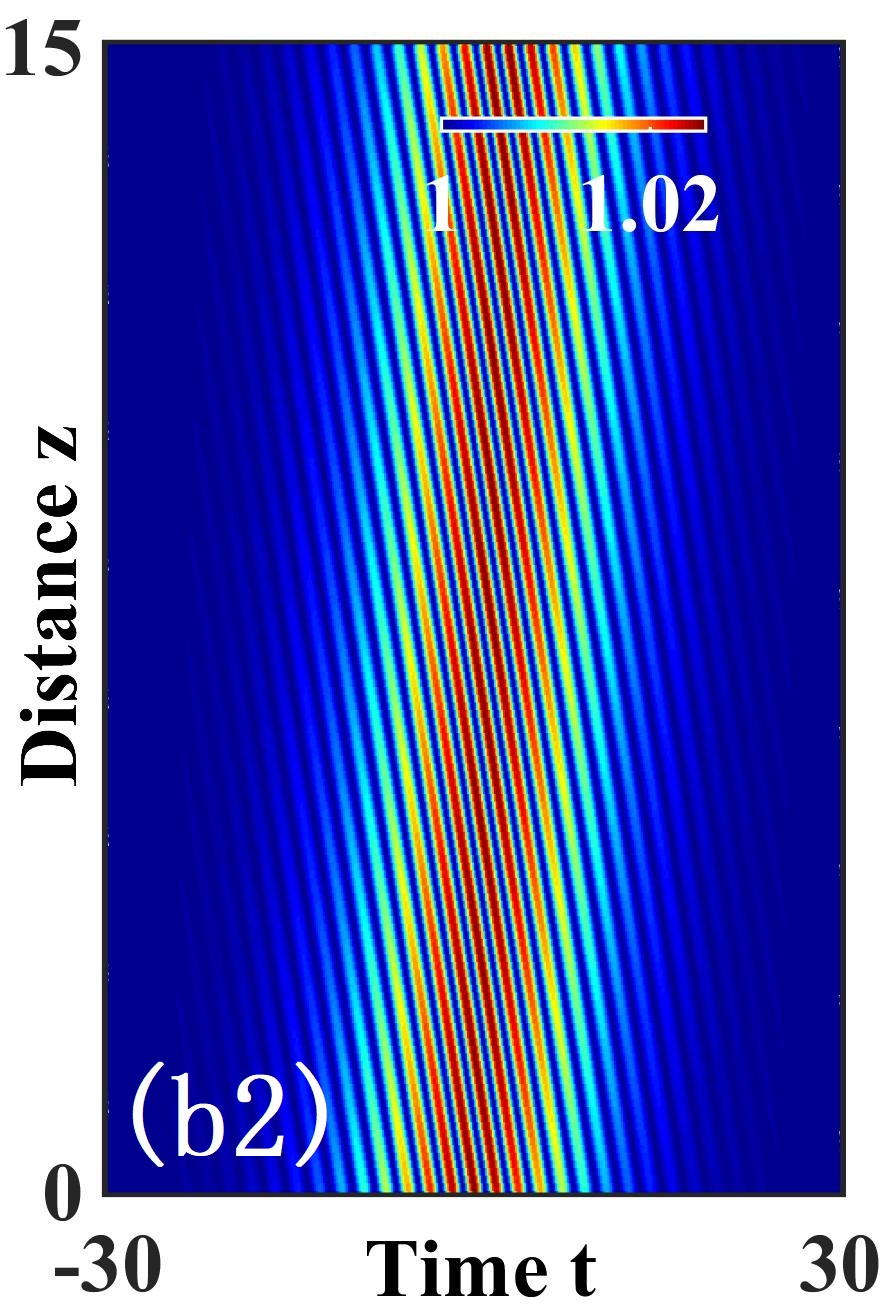}}
\subfigure{\includegraphics[height=31mm,width=23mm]{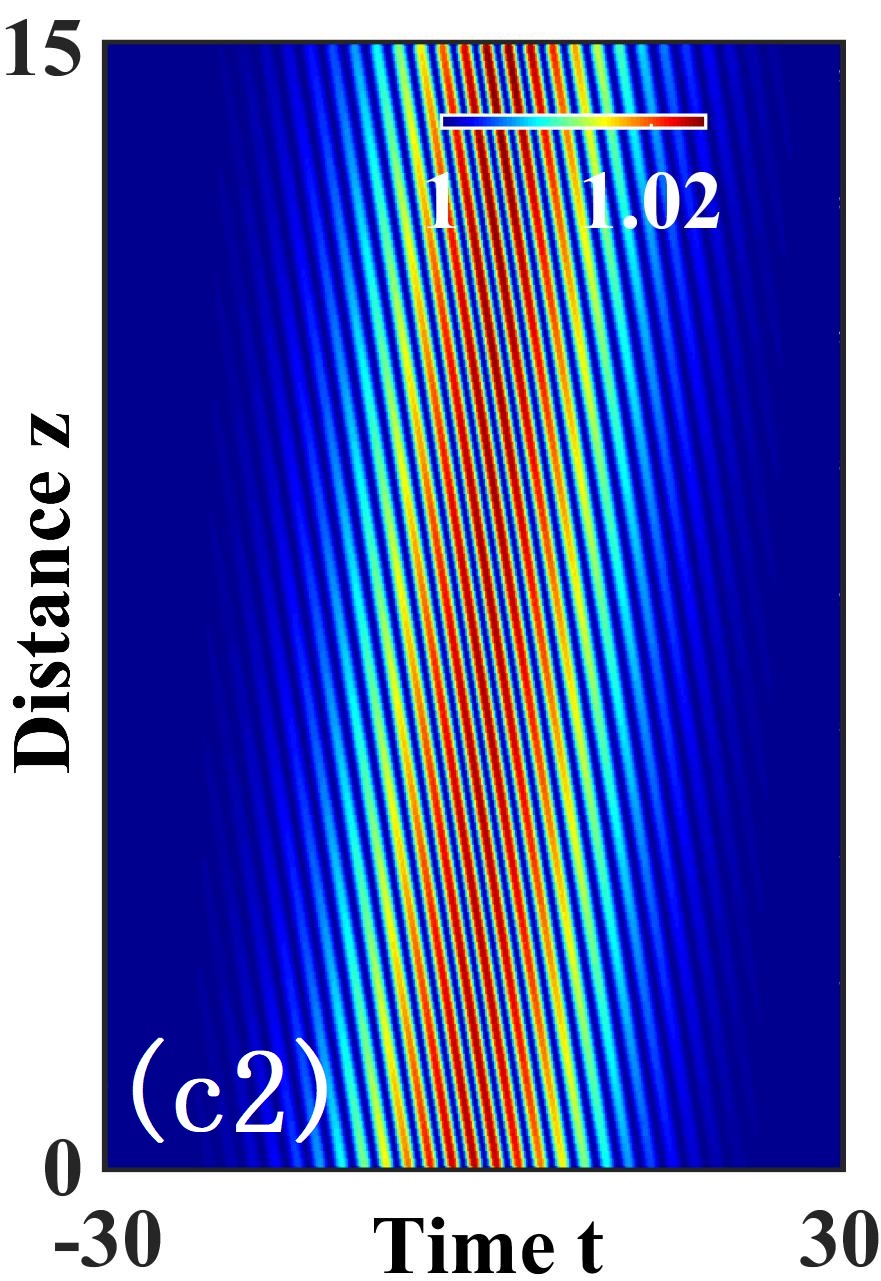}}
\caption{(color online) Numerical confirmation of different SR breathers [(1) half-transition and (2) full-suppression SR modes] from (a) sech pulse, (b) Lorentzian pulse, (c) Gaussian pulse in Fig. \ref{figns}.
The setup is the same as in Fig. \ref{figns}.
}\label{figns1}
\end{figure}

As shown in Fig. \ref{figns}, three different types of nonideal initial states exhibit a consistency compared with the
exact one, despite a slight difference is inevitable. Note also that the Gaussian nonideal type is wider than the other two types.
The interesting finding in Fig. \ref{figns1} is that, different SR breather dynamics can be reproduced excellently by the nonlinear evolution from these
nonideal initial states, although the higher-order effects are considered.  In particular, via a special comparison between them one can readily find that,
the SR breathers generation from the Gaussian initial state
show a wider transverse distribution than those development from the other two initial states. However,
the striking dynamics of SR breathers is observed perfectly.
We note that the smaller and broader that the initial perturbation is (i.e., $\varepsilon\rightarrow0$), the higher that the accuracy of
numerical simulations becomes.
These results confirm the robustness of the SR breather dynamics in the femtosecond regime and will broaden greatly the applicability
of the exact solutions of SR breathers.


\section{Mechanism interpretation by linear stability analysis}
\begin{figure}[htb]
\centering
\subfigure{\includegraphics[height=31mm,width=43mm]{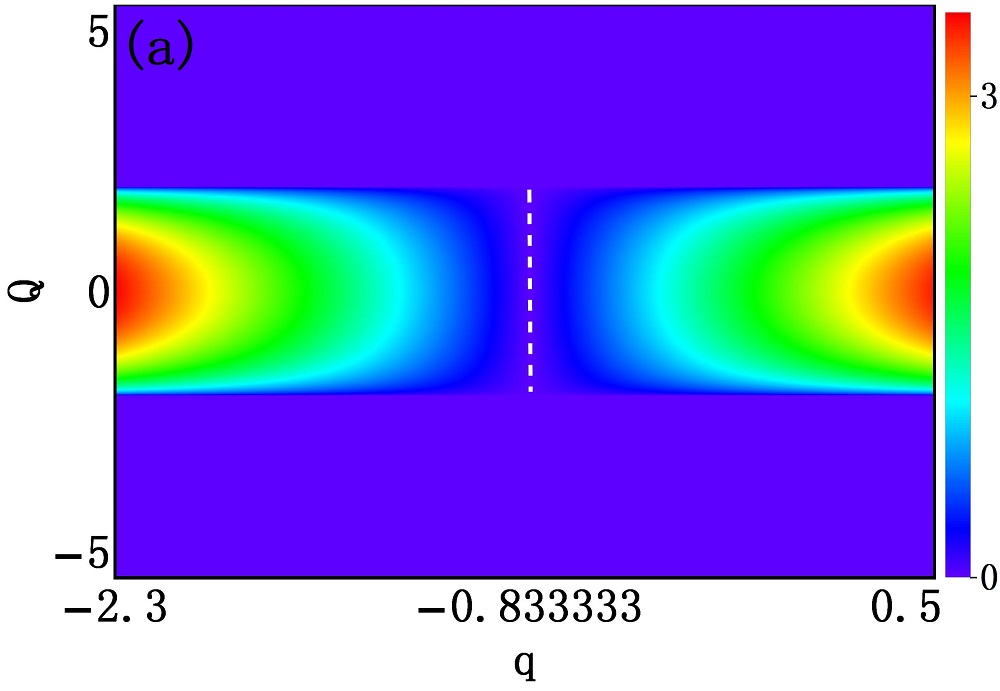}}
\subfigure{\includegraphics[height=31mm,width=42mm]{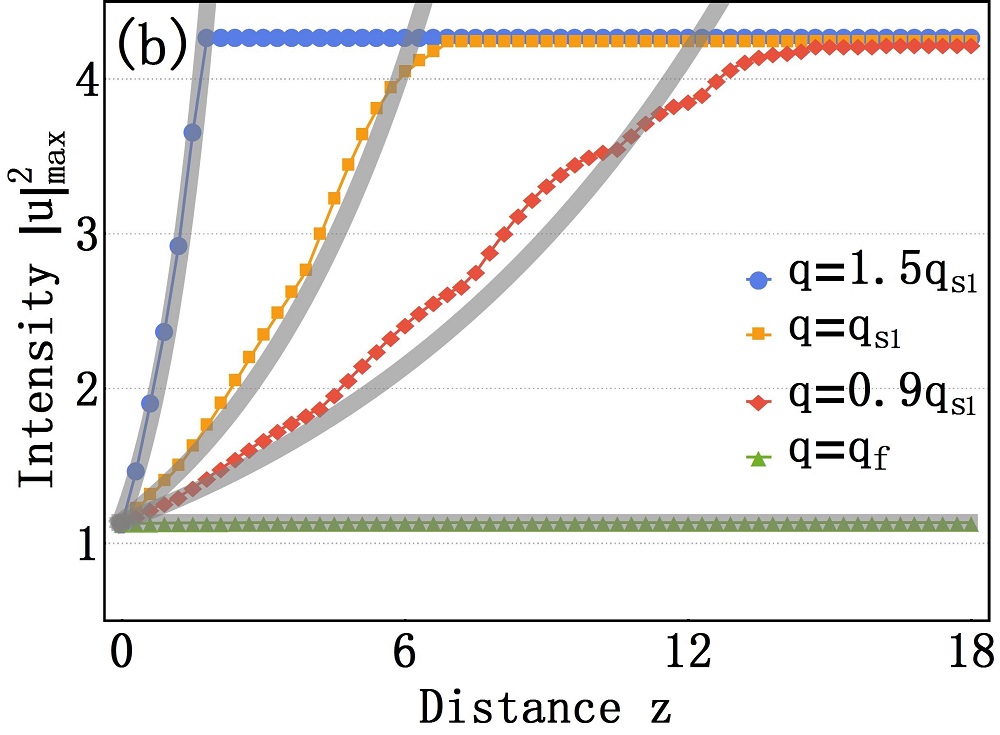}}
\caption{(color online) (a) Distribution of linear MI growth rate $G=b|\textrm{Im}\{\omega\}|$ on $(Q,q)$ plane with $a=1$, $b=2$, (b) Comparison of amplification processes between exact SR solution, Eq. (\ref{equsr}), in Fig. \ref{fig4} (dotted lines) and the linear MI, Eq. (\ref{MI}) (gray solid lines). The dashed line in (a) represents the marginal stability condition: $q=q_f$.
}\label{fig6}
\end{figure}

To understand the formation mechanism of SR modes reported above, our attention is then focused on
the linear stability analysis of the background wave $u_0$ via adding small amplitude perturbed Fourier modes $p$, i.e., $u_{p}=[a+p]e^{i\theta}$, where
$p=f_+e^{i(Q t+\omega z)}+f_{-}^{*}e^{-i(Q t+\omega^* z)}$ with small amplitudes $f_+$, $f_{-}^{*}$, perturbed frequency $Q$, and wavenumber $\omega$.
A substitution of the perturbed
solution $u_p$ into Eq. (\ref{equ1}), followed by the linearization process described in Ref. \cite{MIr2}, yields the
dispersion relation
\begin{equation}
\omega=2(6a^2\beta-q-3q^2\beta-Q^2\beta)\pm\sqrt{Q^2-4a^2}|1-\frac{q}{q_{f}}|.\label{MI}
\end{equation}
MI exists when $\textrm{Im}\{\omega\}<0$ (thus MI exists in the region $|Q|<2a$ with $q\neq q_f$), and is described by the growth rate $G=b|\textrm{Im}\{\omega\}|$, where $b(>0)$ is a real number [see Fig. \ref{fig6}(a)].
Namely, small-amplitude perturbations in this case suffer MI and grow exponentially like $\exp(Gz)$ at the expense of pump waves.

In the previous section we have shown that the initial small-amplitude perturbation $\delta u$ can be amplified exponentially, and eventually becomes the pair of quasi-Akhmediev modes or a mix of quasi-Akhmediev and quasi-periodic waves when $q\neq q_f$.
Note that $\delta u$ possesses a perturbed frequency $(Q=2a\sin\alpha)$ which belongs to the MI region $|Q|<2a$.
We remark that the condition for the MI to exist ($|Q|<2a$, $q\neq q_f$) coincides with that of the nonlinear SR modes where the initial perturbation gets amplified ($Q=2a\sin\alpha$, $q\neq q_f$).

Instead, MI is absent when $|Q|\geq2a$, and $|Q|<2a$ with $q=q_f$ [see Fig. \ref{fig6}(a)]. In this case, the growth rate is vanishing ($G=0$), which implies that the small-amplitude perturbation \textit{cannot} be amplified.
Also, we have shown that $\delta u$ can be suppressed completely when $q=q_f$.
We remark that the marginal stability condition ($|Q|<2a$, $q=q_f$) is consistent with the full-suppression condition of SR modes ($Q=2a\sin\alpha$, $q=q_f$).

For a better understanding, we
compare and analyze the amplification processes of maximum intensity of the initial perturbation obtained from the nonlinear exact SR solution, Eq. (\ref{equsr}) and linear MI, Eq. (\ref{MI}), respectively.
As shown in Fig. \ref{fig6}(b), as $q\rightarrow q_f$,
the amplification evolutions from the exact SR solution and the linear MI growth rate are coincident.
Note that the fluctuation of the maximum intensity appearing in the case $q=0.9q_{s1}$ stems from the beating effects of two waves with a smaller group velocity difference.
Thus, our results reported above show that the characteristics of SR modes and linear stability analysis turns out to be compatible.


We remark that SR breathers, together with well-known Peregrine, and close to Peregrine solutions \cite{na1,na2,na3,MI1} cover all possible discrete spectrum scenarios of MI development from localized small perturbations. The behaviour of continuous spectrum solutions was described in \cite{Biondini}. The question about what scenario is more general still needs further studies.

\section{Conclusions}
We have investigated, analytically and numerically, the existence, characteristic, and mechanism of femtosecond optical SR modes in a fiber which is governed by the higher-order NLSE.
We show, at the first step, the complexity of fundamental modes via a concise phase diagram extracted from a multiparametric analytic solution.
Then the characteristic of SR modes formed by a pair of fundamental waves is revealed.
Such SR waves can exhibit unique half-transition and full-suppression states, which do not have any counterpart in the picosecond regime governed by the standard NLSE. Numerical simulations confirmed that these nontrivial SR states can also be excited from various nonideal initial pulses in both integrable and nonintegrable cases. 

\section*{ACKNOWLEDGEMENTS}
C. Liu appreciates A. Gelash for his valuable comments to improve the paper. This work has been supported by the National Natural Science Foundation of China (NSFC)(Grant Nos. 11705145, 11475135, 11547302, 11434013, 11425522, 11305060), the Fundamental Research Funds of the Central Universities (Grant No. 2015ZD16), the Scientific Research Program Funded by Shaanxi Provincial Education Department (Grant No.17JK0767), and the Major Basic Research Program of Natural Science of Shaanxi Province (Grant Nos. 2017KCT-12, 2017ZDJC-32).


\end{document}